\documentclass[12pt]{article}
\usepackage[margin=1in,tmargin=1.35in,bmargin=1.35in]{geometry} 
\usepackage{amsmath,mathrsfs,amsfonts,amsthm,amssymb}
\usepackage{bm}
\usepackage[round]{natbib}
\usepackage{graphicx}
\usepackage{enumerate}
\usepackage{url} 
\usepackage{threeparttable,booktabs}
\usepackage{multirow}
\usepackage{longtable}
\usepackage{float}
\usepackage{multirow}
\usepackage{subfigure} 
\usepackage{hyperref}
\usepackage{color}
\usepackage[table]{xcolor}

\usepackage{pdflscape}
\usepackage{comment}

\usepackage{xr}
\usepackage{setspace}
\newcommand{\mcal}{\mathcal{M}}

\newcommand{\mcalhat}{\widehat{\mathcal{M}}}
\newcommand{\mhat}{\widehat{m}}

\newtheorem{definition}{Definition}

\newtheorem{theorem}{Theorem}

\usepackage[T1]{fontenc}
\usepackage[utf8]{inputenc}
\usepackage{authblk}

\begin{document}

\title{Model Confidence Bounds for Variable Selection}

\author[1,2]{Yang Li}
\author[3]{Yuetian Luo}
\author[4]{Davide Ferrari}
\author[5,6]{Xiaonan Hu}
\author[7]{Yichen Qin\footnote{corresponding author: yichen.qin@uc.edu}}
\affil[1]{School of Statistics, Renmin University of China}
\affil[2]{Center for Applied Statistics, Renmin University of China}
\affil[3]{Department of Statistics, University of Wisconsin, Madison}
\affil[4]{Faculty of Economics and Management, Free University of Bozen-Bolzano}
\affil[5]{Department of Biostatistics, Yale University}
\affil[6]{School of Mathematical Sciences, University of Chinese Academy of Sciences}
\affil[7]{Department of Operations, Business Analytics, and Information Systems, University of Cincinnati}

\date{}
\maketitle

\begin{abstract}
In this article, we introduce the concept of model confidence bounds (MCB) for 
variable selection in the context of nested models.
Similarly to the endpoints in the familiar confidence interval for parameter 
estimation, the MCB identifies two nested models (upper and lower confidence 
bound models) containing the true model at a given level of confidence. 
Instead of trusting a single selected model obtained from a given model 
selection method, the MCB proposes a group of nested models as candidates and 
the MCB's width and composition enable the practitioner to assess the 
overall model selection uncertainty.
A new graphical tool --- the model uncertainty curve (MUC) --- is 
introduced to visualize the variability of model selection and to compare 
different model selection procedures. 
The MCB methodology is implemented by a fast bootstrap algorithm 
that is shown to yield the correct asymptotic coverage under rather general 
conditions. Our Monte Carlo simulations and real data 
examples confirm the validity and illustrate the advantages of the proposed 
method.

\end{abstract}

\noindent%

\section{Introduction}
\label{sec:intro}

Variable selection is an important and well-studied topic.
Modern analysis is often aimed at selecting a subset of variables from a large number of predictors while attempting to attenuate possible modeling bias.
In the context of linear models, a wealth of methods have been introduced to enhance predictability and to select significant predictors.  
These include popular sparsity-inducing penalization methods such as 
Lasso \protect\citep{tibshirani1996regression}, 
SCAD \protect\citep{Fan2001Variable}, 
Adaptive Lasso \protect\citep{Zou2006The} and so on.  
For an overview of existing methods, please see \protect\citet{Fan2010}.

Regardless of the selection procedure used, variable selection uncertainty is 
an important and ubiquitous aspect of the model selection 
activity. Often using the same variable selection method on different samples from a common population results
in different models. Even for a single sample, different variable selection 
methods tend to select different sets of variables in the presence of 
pronounced noise. 
Motivated by the need to address this model ambiguity, there has been  growing 
interest in developing  model confidence set (MCS) methods, 
which may be broadly regarded as frequentist approaches for obtaining a set of 
models statistically equivalent to a true model at a certain level of 
confidence 
$100(1-\alpha)\%$.  An MCS extends the familiar notion of confidence intervals 
to the model selection framework and enables one to assess the 
uncertainty  associated with a given selection procedure. If the data  are informative, the MCS  contains only a few models 
(exactly one model in the case of overwhelming information), while uninformative data correspond to a large MCS.

\protect\cite{hansen2011model} propose constructing an MCS  from a given 
set of candidate models by a sequence of equivalence tests on the currently remaining models,  followed
by an elimination rule to remove the worst model. Their method obtains a subset of the original models that is meant to contain (or equal) the set of models 
with the best performance under some 
given loss function. \protect\cite{ferrari2015confidence} introduce the notion of  variable selection confidence set (VSCS) for linear regression.
While sharing the same motivation with \protect\cite{hansen2011model}, their method constructs the MCS by a sequence of F-tests and achieves exact coverage 
probability for the true model without necessarily relying on a user-defined 
initial list of models.   They show that,  without  restrictions 
on the model structure (e.g., sparsity), the size of the VSCS is potentially large, thus reflecting the possible model selection uncertainty.  To address this 
issue, they introduce the notion of lower bound models (LBMs) ---  i.e.,  the most parsimonious models that are not statistically significantly inferior to the 
full model at a given confidence level --- and study their properties. 
Previously, 
\protect\cite{shimodaira1998application} advocates the use of a set of models that 
have AIC values close to the smallest among the candidates based on hypothesis testing. \protect\cite{hansen2003choosing, hansen2005model} apply an MCS procedure in 
the context of volatility and forecasting models. \protect\cite{samuels2013forecasting} 
use  the MCS  to select a subset of models  prior to averaging the resulting 
forecasts.

The methods above yield MCS satisfying a nominal coverage 
probability for the true model. 
However, the models contained in the MCS are not 
constrained in terms of their structures, meaning that models in the MCS may  
be drastically different in their compositions with no common 
variables. 
This poses challenges in interpreting the models in the MCS.  
In this paper, we introduce a new 
procedure that computes the so-called model confidence bounds (MCB). 
The MCB is constructed by finding a large and a small model --- 
called upper bound model (UBM) and lower bound model (LBM), respectively --- 
that are nested and where the true model is included between the two
at user-specified confidence level $100(1-\alpha)\%$. 
The LBM and UBM have a rather natural interpretation: The LBM is 
regarded as the most parsimonious model containing indispensable predictors, 
whereas models containing variables beyond 
the UBM include superfluous predictors.  
Note that, even though MCS is more general than MCB since it can be applied to non-nested models, these two methods focus on different objectives.
MCS aims at prediction accuracy, while MCB focuses on model selection uncertainty and on providing more information and interpretation of the model selection results.  It is the similar idea that a confidence interval provides more information than a point estimate.

Our methodology provides a platform for assessing the uncertainty 
associated with different model selection methods.  
Just like using the width of the familiar confidence interval to compare 
different estimator's uncertainty, the practitioner can decide which model 
selection method yields more stable results through comparing the widths and 
compositions of MCB of different methods.
The MCB can also be used as a model selection diagnostic tool. 
If a proposed model is not within the MCB at a certain confidence 
level, there is a 
strong reason to doubt the soundness of its predictors. 
The proposed method is based on bootstrap and may be extended to a wide range of model families.
Finally, to calculate the MCB, we first propose an exact but computationally 
intensive algorithm; we further propose a much more efficient approximated 
algorithm whose performance is found to be comparable to that of the exact 
algorithm. 

The rest of the paper is organized as follows. In Section \ref{sec:meth}, we 
introduce the model confidence bounds (MCB), study its properties, and develop a graphical tool, model uncertainty curve (MUC), to assess model selection uncertainty. 
In Section \ref{sec:algorithms}, we propose fast algorithms to find MCB and study their computational advantages.  
We apply the proposed method to a real data set in Section \ref{sec:real}, and carry out Monte Carlo experiments in Section \ref{sec:simu}. 
We give final remarks in Section \ref{sec:conc} and relegate the 
proofs and additional numerical studies to Web Appendices.

\section{Methodology}
\label{sec:meth}

\subsection{Preliminaries}\label{sec:prelim}
 
We focus on linear regression models. 
Let $\boldsymbol{Y} = (y_1,...,y_n)^T$ be an $n \times 1$ response vector with mean $\bm{\mu}$.  
Suppose
$\boldsymbol{Y} = \boldsymbol{X} \bm{ \theta} + \bm{\epsilon}$
where $\boldsymbol{X}=(\bm{x}_1,...,\bm{x}_n)^T$ is an $n \times p$ matrix of predictors with the $i$th row vector 
$\bm{x}_i = (x_{i1},...,x_{ip})^T \in \mathbb{R}^p$,  
$\bm{\theta}  = (\theta_{1},\dots,\theta_{p})^T \in \mathbb{R}^p$ is the parameter vector, and $\bm{\epsilon} = (\epsilon_1,...,\epsilon_n)^T \sim N_n(\bm{0}, \sigma^2 \boldsymbol{I}_n)$. 
We further make sparsity assumption and assume some of the elements in $\bm{\theta}$ are 
zeros, but we do not know which ones. Let $\mhat$ be the index 
set of some predictors so that it defines 
a possible model and $\emptyset \subseteq \mhat \subseteq 
\{1,...,p\}$.
Then, we have the following definition:
\begin{definition}
  Let $m^{\ast}$ be the index set of predictors with non-zero true 
  coefficients, $m^{\ast} = \{j: \theta_{j}\neq 0, j=1,...,p \}$.
  Let $m_\textup{full}$ be the index set of all predictors, 
  $m_{\textup{full}}= 
  \{j:j=1,...,p \}$.
\end{definition}
Therefore, $m^{\ast}$ represents the true model,  and $m_\textup{full}$ 
represents the full model.  
Without the loss of generality, we assume that the first $p^\ast$ coefficients 
($p^\ast \leq p$) in $\bm{\theta}$ are different from zero so that the true 
model is
$m^\ast=\{1,..., p^{\ast}\}$.  
Let $\mcal_{all} = \big\{m: \emptyset \subseteq m \subseteq 
m_\textup{full}\big\}$ denote the set of all of the possible models.  
In certain situations, prior information on the model structure enables us to 
restrict further the set of all possible models $\mcal_{all}$. For 
example, in polynomial regression, one would include a certain power of the predictor only if all lower-order terms appeared as well. Another case is when certain 
predictors are always protected in the sense that they appear in all candidate models.

Because $\mhat$ represents a subset of all predictors, it can be obtained from 
the variable selection procedure.
Here we focus on penalized likelihood selection methods.  
Specifically, $\mhat= \{j: 
\widehat{\theta}_j \neq 0 , 1 \leq j \leq p \}$ where the estimator 
$\widehat{\bm{\theta}}$ minimizes a penalized likelihood criterion with the form $\ell_n(\bm{\theta})  =  - 2\log L_n(\bm{\theta}; \boldsymbol{Y}, \boldsymbol{X}) + \lambda_n R(\bm{\theta})$ where $L_n(\bm{\theta}; \boldsymbol{Y}, \boldsymbol{X})$ is the likelihood function, $\lambda_n \geq 0$ 
is a user-defined regularization parameter and $R(\bm{\theta})$ is some 
penalty function $R: \mathbb{R}^p \mapsto \mathbb{R}^+$. Throughout the 
paper, $R(\boldsymbol{\theta})$ will be a type of norm. For example, 
$R(\bm\theta) 
=\sum_{j=1}^p  I(\theta_j \neq 0)$ corresponds to the $L_0$-norm and yields a 
number of information theoretical selection criteria, including the Akaike 
information criterion (AIC) for $\lambda = 2$,  and  the Bayesian information 
criterion (BIC) for $\lambda = \log(n)$. Setting  $R(\bm\theta) 
=\sum_{j=1}^p  |\theta_j|$ gives the $L_1$-norm, which corresponds to Lasso.

\subsection{Model Confidence Bounds}\label{sec:MCB_def}

For a given sample $D=\{ (y_i, \bm{x}_i), i =1,\dots, n\}$,  we want to 
find a small model  $\mhat_L =  \mhat_L(D)$ and  a large model  $\mhat_U =  
\mhat_U(D)$ such that the true model $m^\ast$ is nested between  $\mhat_L$ 
and  $\mhat_U$ with a probability at least $1-\alpha$ and $\mhat_L 
\subseteq \mhat_U$.

\begin{definition}[Model Confidence Bounds and Set] \label{def:MCB} The 
$100(1-\alpha)\%$-model 
confidence bounds are defined by the pair of models $\{\mhat_L, 
\mhat_U\}$ such that
\begin{equation}\label{eq:mci}
P\left( \mhat_L \subseteq m^\ast \subseteq \mhat_U \right) \geq 1-\alpha. 
\end{equation}
The models $\mhat_L$ and $\mhat_U$ are called the
lower bound model (LBM)  and the upper  bound model 
(UBM), respectively. The $100(1-\alpha)\%$-model confidence set (MCS) is 
defined as $\mcalhat_\alpha = \{ \mhat:  \mhat_L \subseteq \mhat  
\subseteq 
\mhat_U  \}$.
If (\ref{eq:mci}) is valid as $n \rightarrow \infty$,   $\{\mhat_L, \mhat_U\}$ 
define asymptotic model confidence bounds (AMCB) and asymptotic model 
confidence set (AMCS).
\end{definition}

The above definition extends the usual notion of 
the confidence interval for parameter estimation to the variable selection 
setting. Similarly to  the familiar confidence interval 
for a population parameter,  MCB covers the true model $m^\ast$ with a 
certain probability $1-\alpha$. 
A model smaller than  $\mhat_L$  is regarded as too parsimonious in the sense that it is likely to miss at least 
one important variable, while models with the variables in $\mhat_U$ plus other predictors are considered to be overfitting.  
Similarly to the familiar confidence interval, one can 
obtain a one-sided $100(1-\alpha)\%$-MCB by 
setting $\mhat_L = \emptyset$ or $\mhat_U = m_{\textup{full}}$. 

The pair of models  $\{\mhat_L, \mhat_U\}$ represent two extreme cases, i.e.,  
the most parsimonious and complex models.  
Using these two models (i.e., MCB), we can list all possible 
models nested between those two extremes (i.e., MCS), resulting in an  
easy-to-interpret hierarchical structure. 
Moreover, the difference between the $\mhat_L$ and $\mhat_U$ reflects the model 
selection uncertainty in a given sample. 
When the amount of information in the data is very large, $\mhat_L$ and 
$\mhat_U$ are very similar, and there are only a few models nested between 
MCB. 
In the extreme case of overwhelming information,  we have $\mhat_L =\mhat_U = m^*$ 
and the MCB contains only the true model. 
In most practical situations, we have $\mhat_L \subset \mhat_U$, with the 
discrepancy between $\mhat_L$ and $\mhat_U$ becoming large when the data are 
uninformative.

The size of MCB can be measured by the its width which is defined below.

\begin{definition}[MCB Width]
Let $w(\mhat_L, \mhat_U ) = |\mhat_U | - |\mhat_L|$ be the width associated 
with the model confidence bounds $\{\mhat_L, \mhat_U\}$, where $|\mathcal{A}|$ 
represents the cardinality of set $\mathcal{A}$.
\end{definition}

Note that there are usually multiple MCBs satisfying Equation \eqref{eq:mci}.
Therefore, for simplicity, for any given confidence level $1-
\alpha$, we select the MCB which has the shortest width 
among MCBs satisfying Equation \eqref{eq:mci} and some additional restriction.
We illustrate the selection procedure in details in the following sections.

\subsection{Bootstrap Construction of $100(1-\alpha)\%$-MCB}\label{sec:boot_MCB}

Given the data set $D$, we generate $B$ bootstrap samples $D^{(b)}$, $b=1,\dots,B$.  Then we 
obtain the set of bootstrap models  $\mcal_{boot,B}=\{ \mhat^{(b)}, b=1,\dots, 
B\}$ by applying a model selection method to each bootstrap sample $D^{(b)}$.  
For any 
two nested models, $m_1 \subseteq m_2$ ($m_1$ and $m_2$ denote the index sets 
of some predictors),  it seems quite natural to estimate the probability of the 
event $\{m_1 \subseteq m^\ast \subseteq m_2 \}$ using the following statistic.
\begin{definition}\label{def:BCR} 
The bootstrap coverage rate (BCR) of models $m_1 \subseteq m_2$ is $\widehat{r}(m_1, m_2) =\sum_{b=1}^B I(m_1 \subseteq \mhat^{(b)}  \subseteq m_2 )/B$,
where $\mhat^{(1)}, \dots, \mhat^{(B)}$ are bootstrap models and  $I(\cdot)$ 
is the indicator function.
\end{definition}

For a given confidence level $1-\alpha$, 
to obtain the LBM and UBM,  
we need to find a pair of nested models $m_1 \subseteq m_2$ which satisfy the 
approximate 
inequality $\widehat{r}(m_1, m_2) \geq  1-\alpha$ and have the smallest width 
$|m_2|-|m_1|$. 
To achieve this goal, we first search for the MCB with the highest bootstrap 
coverage rate at different widths, and subsequently form a sequence of MCBs, 
$S$.
Among the sequence of MCBs, we select the final $100(1-\alpha)\%$-MCB to be 
the one having the shortest width while maintaining its bootstrap coverage rate 
greater than or equal to $1-\alpha$.  In other words, we solve the following 
empirical objective function
\begin{equation} \label{eq:approx_program}
(\mhat_L, \mhat_U)= \underset{(m_1, m_2) \in S}{\text{argmin}} 
\left\{ |m_2|-|m_1| : \text{ s.t. } \widehat{r}(m_1, m_2) \geq 1-\alpha 
\right\},
\end{equation}
where $S = \{(m_1^{(i)}, m_2^{(i)}), 0 \leq i \leq p\}$ and $(m_1^{(i)}, 
m_2^{(i)}) = \arg\max_{m_1,m_2}\{\widehat{r}(m_1, 
m_2):\text{ s.t. } |m_2|-|m_1| = i, \emptyset 
\subseteq m_1 \subseteq m_2 \subseteq m_{\textup{full}}\}$.
The set $S$ represents the sequence of MCBs of different widths whose 
bootstrap coverage rates are maximized.
Therefore, $S$ contains the most representative MCBs at each width.
The implementation of this procedure is summarized in Algorithm 1 in 
Section \ref{sec:algorithms} along with a discussion on 
computational complexity and improvement.
It is clear that this approach works as long as the bootstrap coverage rate 
estimates consistently the true coverage rate. A more detailed 
discussion of this issue is deferred to Section \ref{sec:properties}.

\subsection{Assessment of Model Selection Uncertainty}
\label{sec:graphics}

Our proposed MCB can be paired with many model selection methods. 
Therefore, the MCB can be used to assess the uncertainty 
associated with a given model selection method.
Let $\mhat^{(1)},\dots, \mhat^{(B)}$ be bootstrap models under some model 
selection  method. The profiled bootstrap coverage rate (CR) is 
\begin{equation}\label{eq:cr}
\text{CR}(w) = \widehat{r}(\mhat_L, \mhat_U) = 
\sum_{b=1}^{B}I(\mhat_L\subseteq \mhat^{(b)} \subseteq \mhat_U)/B,
 \end{equation}
where $w = |\mhat_{U}|- |\mhat_L|$ is the MCB width.  
Therefore, we treat the CR statistic as a function of the MCB's width $w$. 
For an MCB of width $w$, one would like to use $P(\mhat_L \subseteq 
m^\ast \subseteq \mhat_U)$ 
as a measure of uncertainty for a variable selection method. 
However, in practice the exact probability is unknown, and instead, the CR 
statistic 
is used to approximate $P(\mhat_L \subseteq m^\ast \subseteq \mhat_U)$. 
When a consistent model selection method is used (e.g., BIC, Adaptive Lasso, 
MCP, 
SCAD),  the $P(\mhat_L \subseteq m^\ast \subseteq \mhat_U)$ and the CR are 
typically very close. 

Clearly, a good model selection method would tend to return an MCB with a lower 
width at a given coverage, or a larger coverage value at a given width. 
Thus, we propose to assess the uncertainty of a given model-selection mechanism 
by plotting the pairs of $w/p$ and $\text{CR}(w)$ for all the MCBs in $S$, i.e., 
$\mathscr{P}_{MUC} = \big\{\big(w/p, \text{CR}(w)\big), \ 0 \leq w \leq  
p\big\}$.
The resulting plot is called a model uncertainty curve (MUC).
Essentially, MUC is formed by $w/p$ and $\text{CR}(w)$ of the entire sequence 
of MCBs in $S$.
Note that $S$ contains the sequence of MCBs of different widths whose 
bootstrap coverage rates are maximized.

The MUC of a given variable selection method with good performance will tend to arch towards the upper left corner.
The MUC is in some sense  analogous to that of a receiver operating 
characteristic (ROC) curve used to assess binary classifiers. The ideal model 
selection method has $w/p = 0$ and $\text{CR}(0)=1$, i.e.,  no model selection 
uncertainty at all and perfect coverage (top left corner of the plot). 
Moreover, the area under the MUC (AMUC) can be used as a raw measure of 
uncertainty for the variable selection method under examination.  
A larger value of AMUC implies less uncertainty and more stability of the 
corresponding variable selection method. 
Overall, we can decide which method has the best performance 
according to the 
shape of the MUC and the corresponding AMUC.

\begin{figure}
  \centering
  \includegraphics[width=0.5\textwidth]{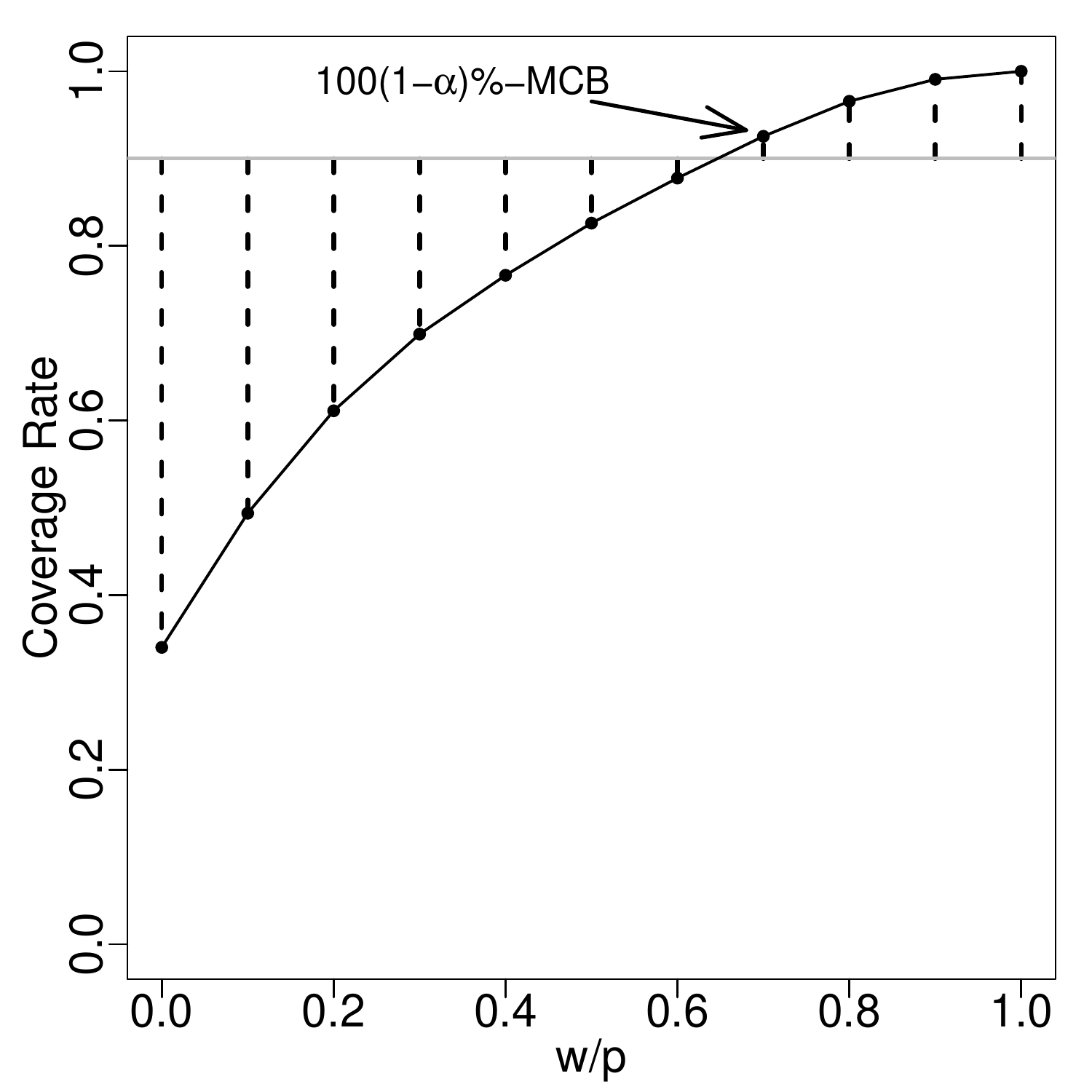}\\
  \caption{Illustration of MUC and $100(1-\alpha)\%$-MCB with $\alpha=0.1$.}\label{fig:MUC_MCB}
\end{figure}

We further present an example of MUC and illustrate its connection to $100(1-\alpha)\%$-MCB.
We simulate a data set according to the linear regression 
$y_i=\sum_{j=1}^{p^\ast} \theta_j x_{ij}+\sum_{j=p^\ast+1}^{p}0 \times x_{ij}  
+\epsilon_i$ with sample size $n=100$, number of predictors $p=10$, number of 
true predictors $p^*=5$, $\theta_j=1$, $\epsilon_i \sim N(0,\sigma^2)$ with 
$\sigma=3$, and 
$\boldsymbol{x}_i 
\sim N_p(\boldsymbol{0},\boldsymbol{I}_p)$.  
We adopt 10-fold cross-validated Adaptive Lasso and plot the MUC in Figure~\ref{fig:MUC_MCB}.
As we can see the MUC arches towards the upper left corner like an ROC curve, which means the model selection method works well for this data set.
In addition, on this MUC curve, each bold dot represents the width and coverage rate of one MCB in $S$.
It is always true that, for that sequence of MCBs, the coverage rate increases monotonically as the width increases.

Suppose we are given the confidence level $1-\alpha$ (i.e., the nominal 
coverage probability), we need to select the $100(1-\alpha)\%$-MCB from $S$
(i.e., MUC).
In Figure~\ref{fig:MUC_MCB}, the confidence level is demonstrated by the gray 
horizontal line.
The coverage probability differences (i.e., empirical coverage probability $-$ nominal coverage probability) of each MCB in $S$ are captured by the vertical dashed lines between the dots and horizontal line.
Clearly, there is a trade-off between the coverage probability difference and 
the width.
However, because of Definition \ref{def:MCB} and Equation 
\ref{eq:approx_program}, only the MCB with nonnegative coverage 
probability difference is eligible to be the $100(1-\alpha)\%$-MCB.  
Therefore, all the dots below the gray line are not eligible.
In addition, since we are searching for the MCB with the shortest width 
according to Equation \ref{eq:approx_program}, the final $100(1-\alpha)\%$-MCB
is the dot closest to and yet above the gray line.
The rest of MCBs above the gray line have larger widths compared to the selected
$100(1-\alpha)\%$-MCB due to the monotonicity of the coverage rate in width.

\subsection{Asymptotic Coverage}\label{sec:properties}

The quality of the model selection method can be measured by the underfitting 
and overfitting probabilities, i.e., the probability of the events $\mcal^{-}_n = \{ \mhat \nsupseteq m^\ast \}$ and $\mcal^{+}_n = \{ \mhat \supsetneqq m^\ast\}$.
Note that $\mcal^{-}_n$ represents some predictors in the 
true model are missed.  
Meanwhile, $\mcal^{+}_n$ represents the predictors in the true model are selected plus some 
additional superfluous terms.  
A model selection procedure is consistent if $P(\mhat = m^\ast) \rightarrow 1$ 
as $n \rightarrow \infty$ for every $\bm{\theta} \in \mathbb{R}^p$, which  
occurs if 
$P(\mcal^{-}_n) \rightarrow 0$ and $P(\mcal^{+}_n) \rightarrow 0$ as $n 
\rightarrow \infty$. If $P(\mcal^{-}_n) \rightarrow 0$ but $\mhat$ is not 
consistent, then we call the procedure conservative.

\begin{theorem}\label{thm:coverage}
Assume: (A.1) (Model selection consistency) $P(\mcal_n^{-}) = o(1)$ and 
$P(\mcal_n^{+}) = o(1)$; (A.2) (Bootstrap validity). For the re-sampled model 
 $\mhat^{(b)}$, assume $P(\mhat^{(b)} \neq \mhat) = o(1)$. Then, for the 
$\mhat_L$ and $\mhat_U$ solving program (\ref{eq:approx_program}), we have $P(\mhat_L \subseteq  m^\ast \subseteq \mhat_U) \geq 1- \alpha + o(1)$.

\end{theorem}

The above theorem shows that if  the model selection procedure is consistent 
(i.e., the probability of overfitting or underfitting the underlying model 
becomes small as the sample size increases), then the CR statistic 
estimates consistently the true coverage probability associated with models 
$m_1 \subseteq m_2$.  
Note that assumptions A.1 and A.2 are common conditions in the  literature.  
For example, \protect\citet{Fan2010} establish the model selection consistency (as part of the oracle property) as a standard property for many methods.  
\protect\citet{Potscher2010} and \protect\citet{Leeb2006} have presented similar assumptions in their results.
Lastly, \protect\citet{Liu2013} and \protect\citet{Chatterjee2013} have also stated the bootstrap validity for their methods.
For each assumption, we list a few examples as follows.

For Assumption A.1 of model selection consistency, many existing model 
selection methods are equipped with such a property.  
In the case of Lasso estimation, \protect\citet{Zhao2006} establish the condition for selection consistency as the Strong Irrepresentable Condition.
This condition can be interpreted as: 
Lasso estimate selects the true model consistently if and (almost) only if the predictors that are not in the true model are ``irrepresentable'' by 
predictors that are in the true model.
\protect\citet{Zhao2006} demonstrate that the Strong Irrepresentable 
Condition holds when the correlation matrix of 
the covariates belongs to constant positive correlation matrix, 
power decay correlation matrix and many others.
In addition, selection consistency has also been established for various penalization methods, such as Adaptive Lasso, SCAD, MCP and LAD.  
\protect\citet{Fan2001Variable} demonstrate that nonconcave penalty such as SCAD 
enables the penalized regression estimator to enjoy the oracle property, 
which means, with probability tending to 1, the estimate correctly identifies 
the true zero and nonzero coefficients.
In addition, the estimate for the true nonzero coefficients works as well as if the correct sub-model were known.  
\protect\citet{Zou2006The} demonstrates Adaptive Lasso is equipped with the selection consistency.
\protect\citet{Zhang2010} and \protect\citet{Wang2007} further show the selection consistency of MCP and LAD.

Lastly, there are other existing model selection methods based on minimizing 
criterion of the form,
$\text{IC}(\mhat)=\log 
\left(\textup{RSS}_{\mhat}/n\right)+|\mhat|C_n/n$,
where $\textup{RSS}_{\mhat}$ represents the residual sum of squares.  
It is well known that the model minimizing $\text{IC}(\mhat)$ is a 
consistent estimate of the true model, 
if the penalty satisfies that $C_n/n \to 0$ and $C_n \to \infty$ as $n \to 
\infty$.
If $C_n$ is bounded, then the corresponding model selection method is 
conservative.
Therefore, it is straightforward to show that the celebrated minimum BIC 
approach is consistent since $C_n =\log n$.
On the other hand, minimum AIC approach is conservative since $C_n = 2$.

For Assumption A.2 of bootstrap validity, it has been established in the 
literature that several existing bootstrap procedures are valid for various 
model selection methods.
For example, residual bootstrap \protect\citep{Freedman1981} provides a valid bootstrap 
approximation to the sample distribution of the least square estimates.
In addition, \protect\citet{Chatterjee2011} show that, using residual bootstrap, 
the sampling distribution of the Adaptive Lasso estimate 
$\widehat{\boldsymbol{\theta}}$ can be 
consistently estimated by the sampling distribution of the bootstrap estimate 
$\widehat{\boldsymbol{\theta}}^{(b)}$.  
In other words, we have $\varrho(H^b,H) \overset{p}{\to} 0$ where 
$\varrho(\cdot,\cdot)$ represents the Prohorov metric on the set of all 
probability measures on $(\mathbb{R}^p , \mathcal{B}(\mathbb{R}^p ))$, 
and $H^b$ and $H$ are the asymptotic distributions of the centered and scaled 
estimates,
$\sqrt{n}(\widehat{\boldsymbol{\theta}} - \boldsymbol{\theta})$ and 
$\sqrt{n}(\widehat{\boldsymbol{\theta}}^{(b)}-\widehat{\boldsymbol{\theta}})$.
Similarly, residual bootstrap is able to produce a valid 
sampling distribution for SCAD, MCP and other studentized estimators \protect\citet{Chatterjee2013}.  
Throughout the article, we mostly adopt residual bootstrap.

On the other hand, \protect\citet{Chatterjee2010} and \protect\citet{Camponovo2015} 
demonstrate that residual bootstrap and pairs bootstrap do not provide a valid 
approximation of the sampling distribution for Lasso estimation.
A modified bootstrap method by \protect\citet{Chatterjee2011} provides a valid 
approximation to the sampling distribution of the Lasso estimator, which 
involves hard-thresholding the Lasso estimate when generating the bootstrap 
samples.  We have adopted this algorithm for all the numerical studies on Lasso 
estimate.  In addition, there is also a modified pairs bootstrap available 
for Lasso estimate \protect\citep{Camponovo2015}.  The details of the residual bootstrap and modified residual bootstrap are described in Web Appendix A.

\section{Algorithms} \label{sec:algorithms}

\subsection{Naive Implementation by Exhaustive Search}\label{sec:algo1}

While traditional confidence intervals for parameter estimation are typically 
computed by finding lower and upper bounds based on a given confidence level, 
our MCB is numerically determined in a reverse way due to computational 
concerns. 
Specifically, at each width $w$, we first search for an 
$\textup{MCB}(w)$ of width $w$ which has the highest bootstrap coverage rate.  
Hence we have a sequence of $\textup{MCB}(w)$ for $w=0,...,p$.
Among these $\textup{MCB}(w)$s, we select the final $100(1-\alpha)\%$-MCB to be 
the one having the shortest width while maintaining its bootstrap coverage rate 
greater than or equal to $1-\alpha$.
This straightforward procedure is detailed in Algorithm 1.

\indent \textbf{Algorithm 1}: Naive $100(1-\alpha)\%$-model confidence bounds

\begin{itemize}\itemsep -2pt
  \item[1.] Generate $B$ bootstrap samples and obtain $B$ bootstrap models 
  $\mhat^{(1)}, \dots, \mhat^{(B)}$.
   \item[2.] For $w=0,\dots,p$, obtain $\textup{MCB}(w)$ of width $w$ by 
   $\{ \mhat_L(w),  \mhat_U(w)\} = 
      \underset{m_1, m_2}{\text{argmax}} \{ \widehat{r}(m_1, m_2): $\\
      $\text{ s.t. } |m_1|-|m_2|=w, m_1 \subseteq m_2 \}.$

  \item[3.] Among the sequence of $\textup{MCB}(w)$, choose the final 
  $100(1-\alpha)\%$-MCB to be $\textup{MCB}(w^*)$, where $
  w^\ast = \min \left\{ w: \text{ s.t. } 
  \widehat{r}(\mhat_L(w), \mhat_U(w))\geq 1-\alpha, 0\leq w \leq p \right\}$.
\end{itemize}
 
We further explore the computational cost of Algorithm 1. 
The number of iterations involved in Step 2 is $\Omega_\text{1} = 
\sum_{w=0}^{p} \Omega_{1}(w)$ where 
$\Omega_\text{1}(w) = \binom{p}{0} \binom{p}{w} + \binom{p}{1}\binom{p-1}{w} + 
 \dots + \binom{p}{p-w}\binom{w}{w}$.
Although Algorithm 1 returns an exact solution, this naive strategy 
essentially 
requires exhaustive enumeration, and is therefore applicable only in the 
case of small $p$.
Next, we turn our interest to another strategy that achieves similar accuracy 
while involving a greatly reduced computational burden.

\subsection{Implementation by Predictor Importance Ranking}\label{sec:algo2}

The theoretical findings in Section \ref{sec:properties} show that the 
predictors in the true model are selected with higher frequencies than 
irrelevant predictors. 
Hence we propose Algorithm 2 based on the fact that the more frequently 
selected the predictor is, the more likely it is in MCB.

Specifically, let the importance of the $j$th predictor be measured by its 
selection frequency among $B$ bootstrap models, $\overline{\pi}_j  = B^{-1} 
\sum_{b=1}^{B} I(j \in \mhat^{(b)})$. 
Let $\Pi= (u_1,\dots, u_p)$ be the arrangement of indices in $\{1,\dots, p\}$ 
induced by the ordered frequencies 
$
\overline{\pi}_{u_1} > \cdots > \overline{\pi}_{u_p}
$
(assuming no ties). 
The ordering $\Pi$ induces a natural ranking of predictors.

When searching for $\text{MCB}(w)$ of width $w$, 
we consider constructing the LBM by taking the $k$ most important 
predictors according to the ordering $\Pi$ and $0 \leq k \leq p-w$, and 
constructing the UBM by adding the next a few 
important predictors until reaching the desired width $w$, i.e., $\mhat_L(k)  = 
\{ u_1, \dots, u_k\}$ and 
$\mhat_U(k,w)  = \{ u_1, \dots, u_k, u_{k+1}, \dots, u_{k+w}\}$. 
Thus $\textup{MCB}(w)$ can be simply and efficiently 
determined by $\{ \mhat_L(k^*), \mhat_U(k^*, w) \}$ where 
$k^* =  \arg\max_{k: 0 \leq k \leq p- w} \widehat{r}(\mhat_L(k), \mhat_U(k,w))$.
By setting $w=0,...,p$, we again have a sequence of $\textup{MCB}(w)$. 
Finally, we choose the $\textup{MCB}(w^*)$ with the shortest width while maintaining its bootstrap coverage rate no smaller than the nominal confidence level 
$100(1-\alpha)\%$ as our final MCB.  

The above discussion leads to our Algorithm 2.  
We call this new algorithm $100(1-\alpha)\%$-MCB construction by predictor 
importance ranking (PI-MCB).

\indent \textbf{Algorithm 2}: $100(1-\alpha)\%$-model confidence bounds by 
predictor importance ranking
\begin{itemize}\itemsep -2pt
   \item[1.] Same as Step 1 of Algorithm 1.
   \item[2.] For $j=1,\dots, p$, obtain predictor importance  $\overline{\pi}_j 
   = B^{-1} \sum_{b=1}^{B} I(j \in \mhat^{(b)})$, and generate the 
   ordering $\Pi = \{u_1, \dots, u_p \}$ induced by $\overline{\pi}_{u_1}> 
   \cdots > \overline{\pi}_{u_p}$.   
   \item[3.] For $w=0, 1,\dots, p$, obtain the $\text{MCB}(w)$ of width 
   $w$ by $\text{MCB}(w) = \arg\max_{m_1,m_2}
         \widehat{r}(m_1, m_2) $  where $m_1= \{u_1,\cdots,u_k \}$ 
         and $m_2=\{u_1,\cdots,u_{k+w}\}$.
   \item[4.] Same as Step 3 of Algorithm 1.
  \end{itemize}

Algorithm 2 (PI-MCB) is extremely fast compared to Algorithm 1. 
The number of iterations of Algorithm 2 is 
$\Omega_{2} = (p+1)(p+2)/2$.
When $p=10$, Algorithms 1 and 2 require about 60,000 and 60 iterations, 
respectively.  
Meanwhile, the number of iterations of Algorithm 1 increases very 
rapidly in $p$, suggesting that this is not a viable algorithm for 
high-dimensional regression problems.  
In addition to the computational advantages, Algorithm 2's performance is also 
justified by the following theorem.

\begin{theorem}\label{thm: alg}
Suppose there are $p$ predictors $X_j,  j=1,\cdots,p$, and $B$ bootstrap  models $\mhat ^{(b)}, b=1,\cdots,B$.
For any given width $w$, 
let $\{\mhat_{L,1},\mhat_{U,1}\}$ and $\{\mhat_{L,2},\mhat_{U,2}\}$ be the MCB 
of width $w$ by Algorithm 1 and Algorithm 2, respectively,
and their CRs are $\widehat{r}(\mhat_{L,1},\mhat_{U,1})$ and  $\widehat{r}(\mhat_{L,2},\mhat_{U,2})$. 
Assume: (A.3) $I(X_j)$'s, $j=1,\cdots,p$, are mutually independent where 
$I(X_j)$ denotes the event that $X_j$ is selected in the model.
Then, as $B \to \infty$, we have $|\widehat{r}(\mhat_{L,1},\mhat_{U,1}) - 
\widehat{r}(\mhat_{L,2},\mhat_{U,2})|=o_p(1)$.

\end{theorem}

The above theorem shows that, when all predictors are mutually independently 
selected, Algorithm 1 and Algorithm 2 yield the same performance in terms of 
coverage rate.
However, in practice, Assumption A.3 is very difficult to satisfy or to verify.
Nevertheless, the theorem provides some key insights for Algorithm 2.  
Through simulation, we have shown that Algorithms 1 and 2 perform very 
similarly even when the predictors are moderately correlated.
We conduct an example to illustrate the connection 
between these Algorithms in Web Appendix A.

\section{Real Data Analysis}
\label{sec:real}

We illustrate the proposed method using the diabetes data set 
\protect\citep{efron2004least} which consists of measurements on $n=442$ diabetic 
patients. There are $p=10$ predictors: body mass index (\texttt{bmi}), 
lamotrigine (\texttt{ltg}), 
mean arterial blood pressure (\texttt{map}), 
total serum cholesterol (\texttt{tc}), 
sex (\texttt{sex}), 
total cholesterol (\texttt{tch}), 
low- and high-density lipoprotein (\texttt{ldl} and \texttt{hdl}), 
glucose (\texttt{glu}) and age (\texttt{age}).
The response variable is a measurement of disease progression one year after 
baseline.
Using such a data set, we construct MCB via Adaptive Lasso, Lasso and 
stepwise regression using BIC and also construct VSCS 
\protect\citep{ferrari2015confidence}.

\begin{figure}
  \centering
  \includegraphics[width=0.5\textwidth]{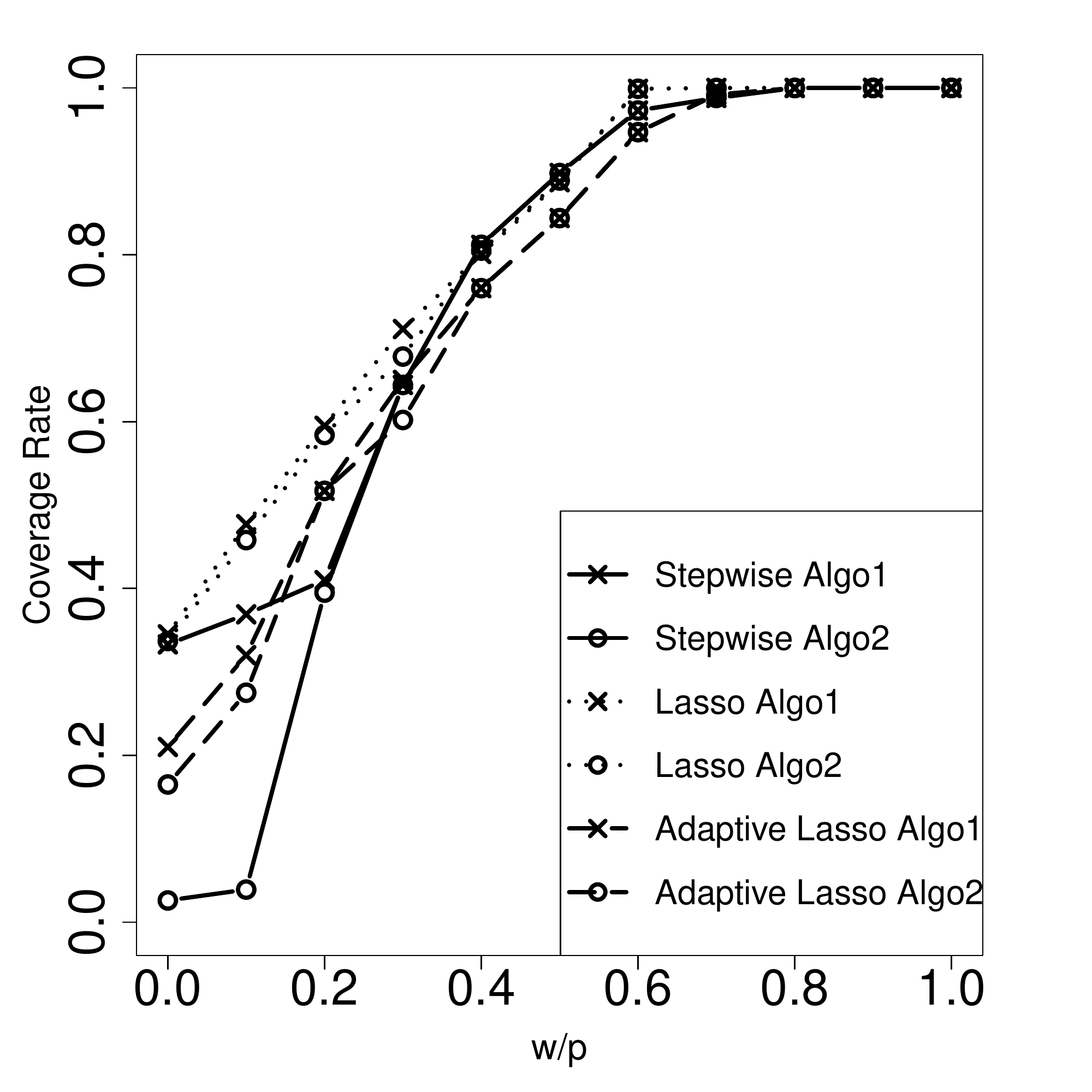}\\
  \caption{MUC of three model selection methods (Adaptive Lasso, Lasso, and 
  stepwise regression using BIC) when applying to the diabetes data set. 
  $B=1000$ bootstrap samples are generated from the original dataset.  
  The tunning parameters of Adaptive Lasso and Lasso are chosen by 10-fold 
  cross-validation.  Two proposed algorithms are applied.}\label{fig:diabetesCR}
\end{figure}

In Figure  \ref{fig:diabetesCR}, we first compare the model selection methods 
using the MUC.
As we can see, the coverage rate increases with the width $w$.
The MUCs of all three methods arch towards the upper left corner.
For Adaptive Lasso's MUC, 
when $w=2$ and $w/p=0.2$, the coverage rate is around 0.4,
meaning the MCB captures only about 40\% of the bootstrap 
models.
When $w>6$, the coverage rate stays above $0.9$, meaning that the MCB contains 
more than 90\% bootstrap models.
Recall that the interpretation of the MUC is similar to that of the more 
familiar ROC curve.

Table \ref{tab:MCB} further shows the 95\%- and 75\%-MCB of different model 
selection methods. 
When the confidence level increases, the LBM becomes smaller while 
the UBM becomes larger, and the cardinality of the MCB (i.e. number of unique 
models nested between LBM and UBM) also increases. 
Note that the bootstrap coverage rate is slightly larger than the confidence level due to 
the design of the algorithms.
We also report the single selected models using these methods in the same table.
As we can see, for each method, the 95\%-MCB always contains the corresponding 
single selected model.
According to our MCB results, the predictors \texttt{bmi, ltg, map} are 
considered most indispensable since they appear in most of the LBMs.  
Meanwhile, \texttt{age} is not included in any UBMs and should be excluded in 
the modeling process.
Such a conclusion is also consistent with other existing studies 
\protect\citep{lindsey2010variable,efron2004least}.

We also apply VSCS on this data set using the same confidence levels, 95\% and 
75\%.
The VSCS contains much more models than MCB (i.e., higher 
cardinality), 
suggesting that MCB can identify the true model more 
efficiently.
For example, VSCS returns 528 and 288 unique models at confidence levels 95\% 
and 75\%, respectively.  
Compared these cardinalities with these of MCB (i.e., 16 and 64 for Adaptive Lasso, 16 and 32 for Lasso), we can 
see the clear advantage of MCB.  
Note that the models returned by MCB are always nested between the LBM and UBM, 
whereas the models returned by VSCS do not have such a structure, as
they are simply the survivors of the F-test.
Therefore, the models in VSCS are scattered in the entire space of all possible 
models (i.e., $2^p$ models).
In addition, VSCS can be ``roughly'' considered as having multiple LBMs 
and having the full model as UBM since the full model by default will survive 
the F-test.
However, note that VSCS does not contain all the models that are nested between its LBMs and UBM.
Therefore, it is much harder to interpret VSCS results.
We have summarized all LBMs and UBMs of VSCS in Table \ref{tab:MCB}.
At the confidence levels 95\% and 75\%, VSCS contains 7 and 4 different LBMs, 
respectively. They are all consistent with the LBM and UBM of MCB, in the sense that the predictors frequently appearing in LBMs of VSCS also appear in LBM of MCB.
The predictors less frequently appearing in LBMs of VSCS are mostly in UBM of MCB.

\section{Simulations}
\label{sec:simu}

We investigate the performance of the proposed method by Monte 
Carlo (MC) experiments. 
Each MC sample is generated from the model, $y_i=\sum_{j=1}^{p^\ast} \theta_j 
x_{ij}+\sum_{j=p^\ast+1}^{p}0 \times x_{ij}  + \epsilon_i$, where 
$p^\ast$ is the number of true variables and $p$ is the number of 
candidate variables.
We simulate the random error according to $\epsilon_i \sim 
N(0,\sigma^2)$ and the covariate according to $\boldsymbol{x}_i \sim 
N_p(\boldsymbol{0},\boldsymbol{\Sigma})$ where 
$\boldsymbol{\Sigma}=[\Sigma_{ij}]_{p \times p}$ and $\Sigma_{ij}=\rho^{|i-j|}$.
Additional simulations are in Web Appendix A.

Note that our simulation setting with the power decay correlation matrix $\boldsymbol{\Sigma}$ satisfies the 
Strong Irrepresentable Condition as proved by Corollary 3 in \protect\citet{Zhao2006}, 
which provides the selection consistency of Lasso estimate.
Other model selection methods used in this section include Adaptive Lasso, SCAD, MCP, LAD Lasso \protect\citep{Wang2007}, and SQRT Lasso \protect\citep{Belloni2011}, which are all proved to hold the property of selection consistency.
Therefore, Assumption A.1 is satisfied.
In addition, the modified residual bootstrap is used for Lasso estimate.
The residual bootstrap is used for Adaptive Lasso, SCAD, MCP, LAD Lasso, and SQRT Lasso. 
These bootstrap algorithms provide valid approximations to the sampling distributions of the estimates and further guarantee  Assumption A.2.

\subsection{Comparison of Algorithms}\label{sec:comp_algo}

\begin{figure}[tbp]
\centering
\subfigure[$p=8$]{
\includegraphics[width=0.3\textwidth]{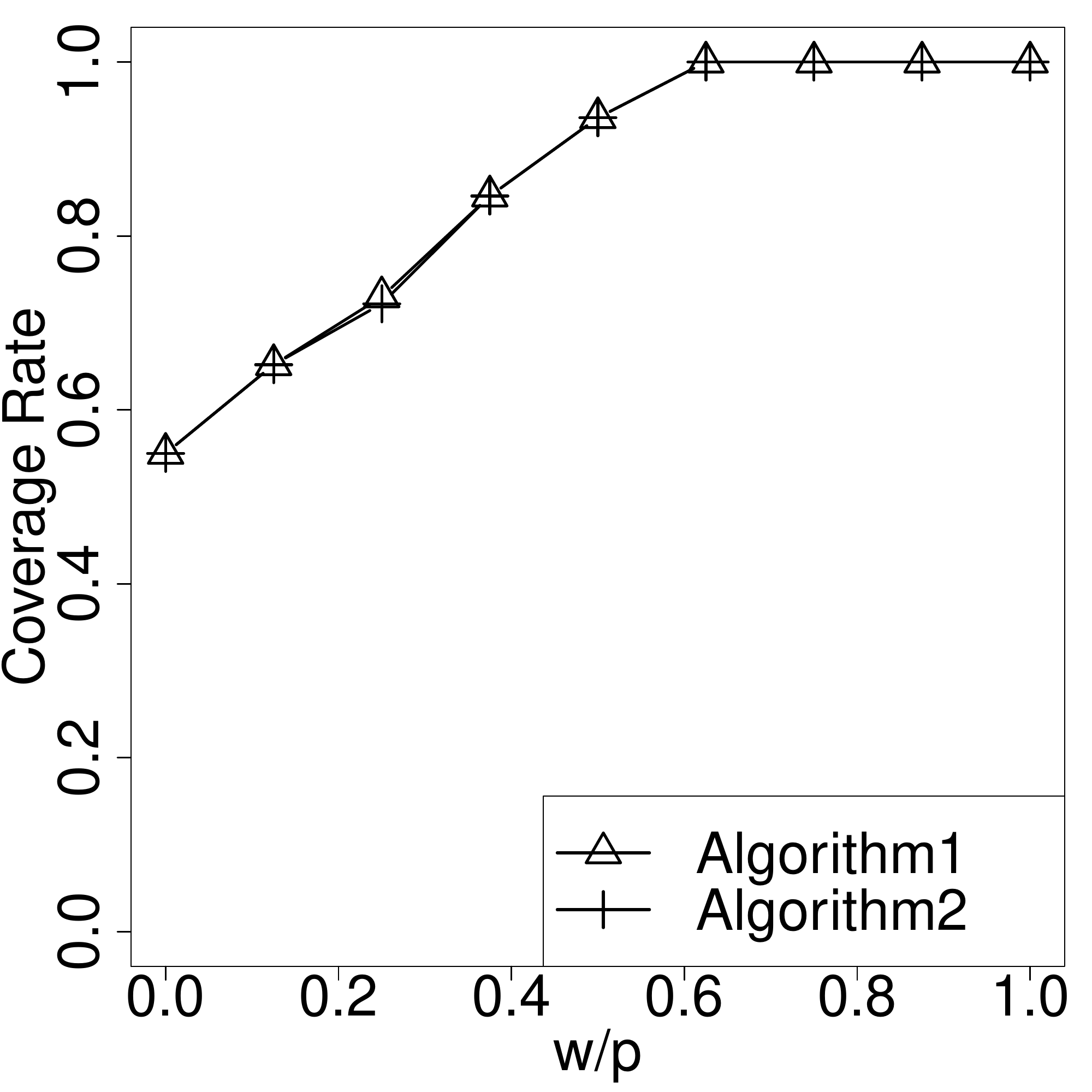}
}
\hspace{0 in} 
\subfigure[$p=10$]{
\includegraphics[width=0.3\textwidth]{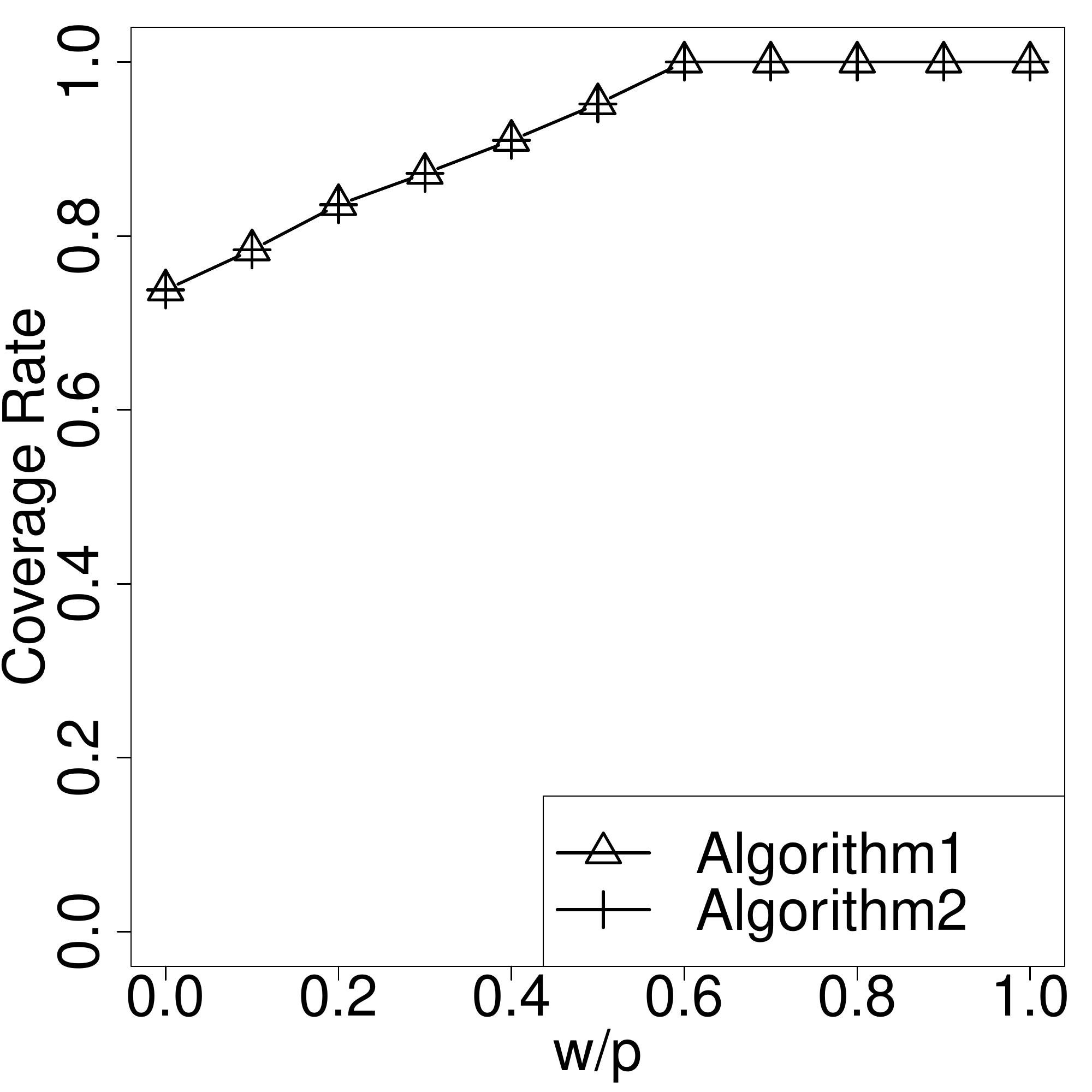}
}
\hspace{0 in} 
\subfigure[$p=15$]{
\includegraphics[width=0.3\textwidth]{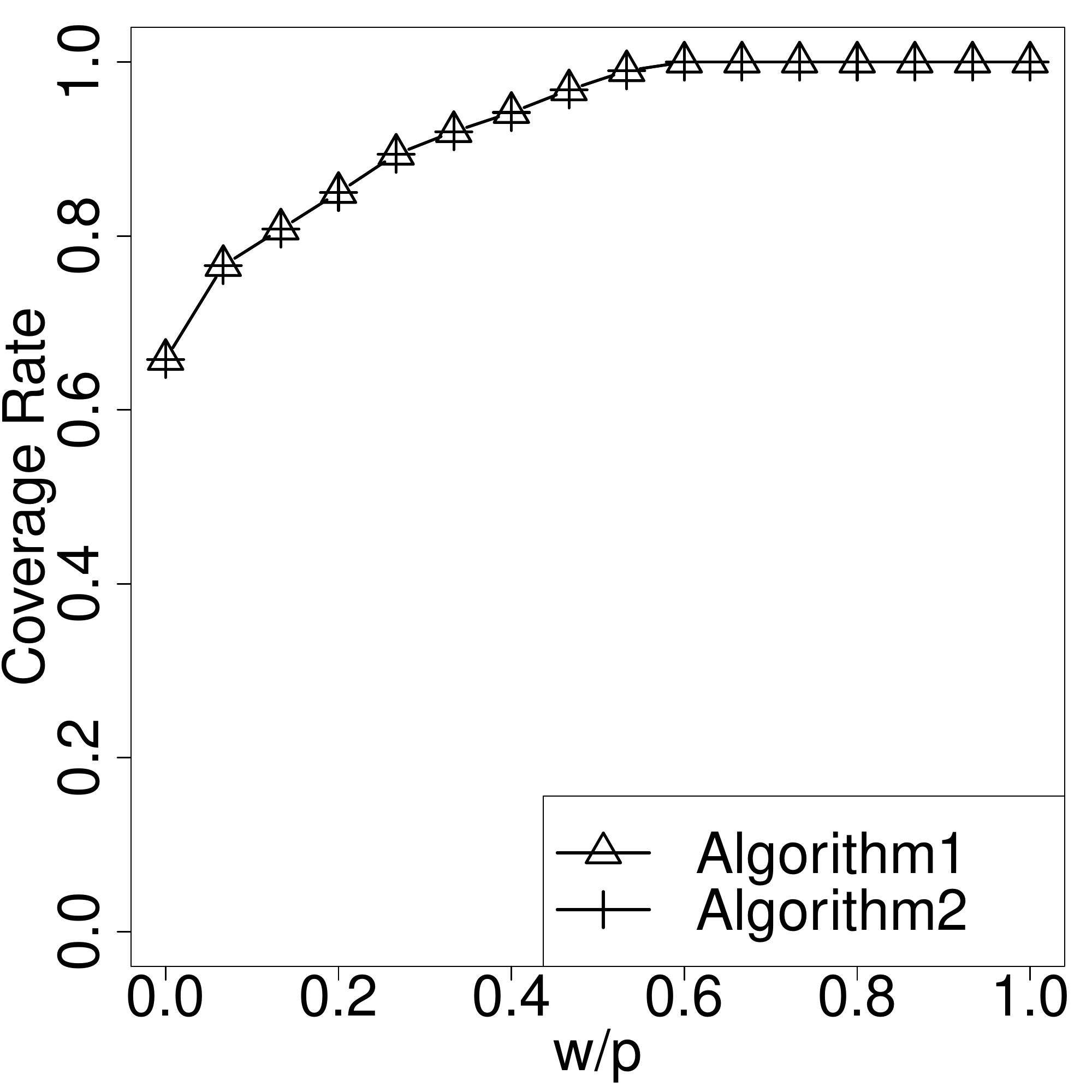}
}
\subfigure[$p=50$]{
\includegraphics[width=0.3\textwidth]{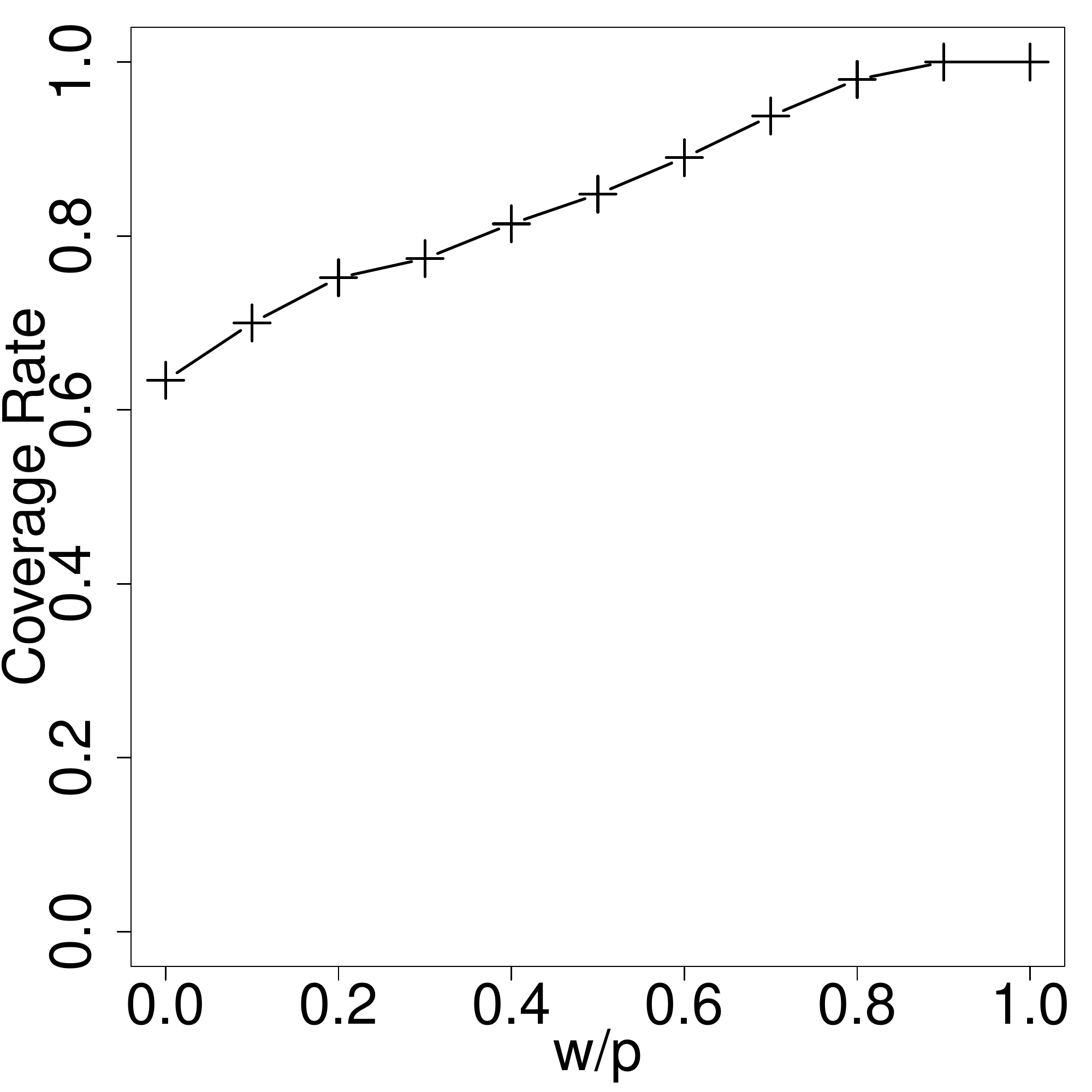}
}
\hspace{0 in} 
\subfigure[$p=100$]{
\includegraphics[width=0.3\textwidth]{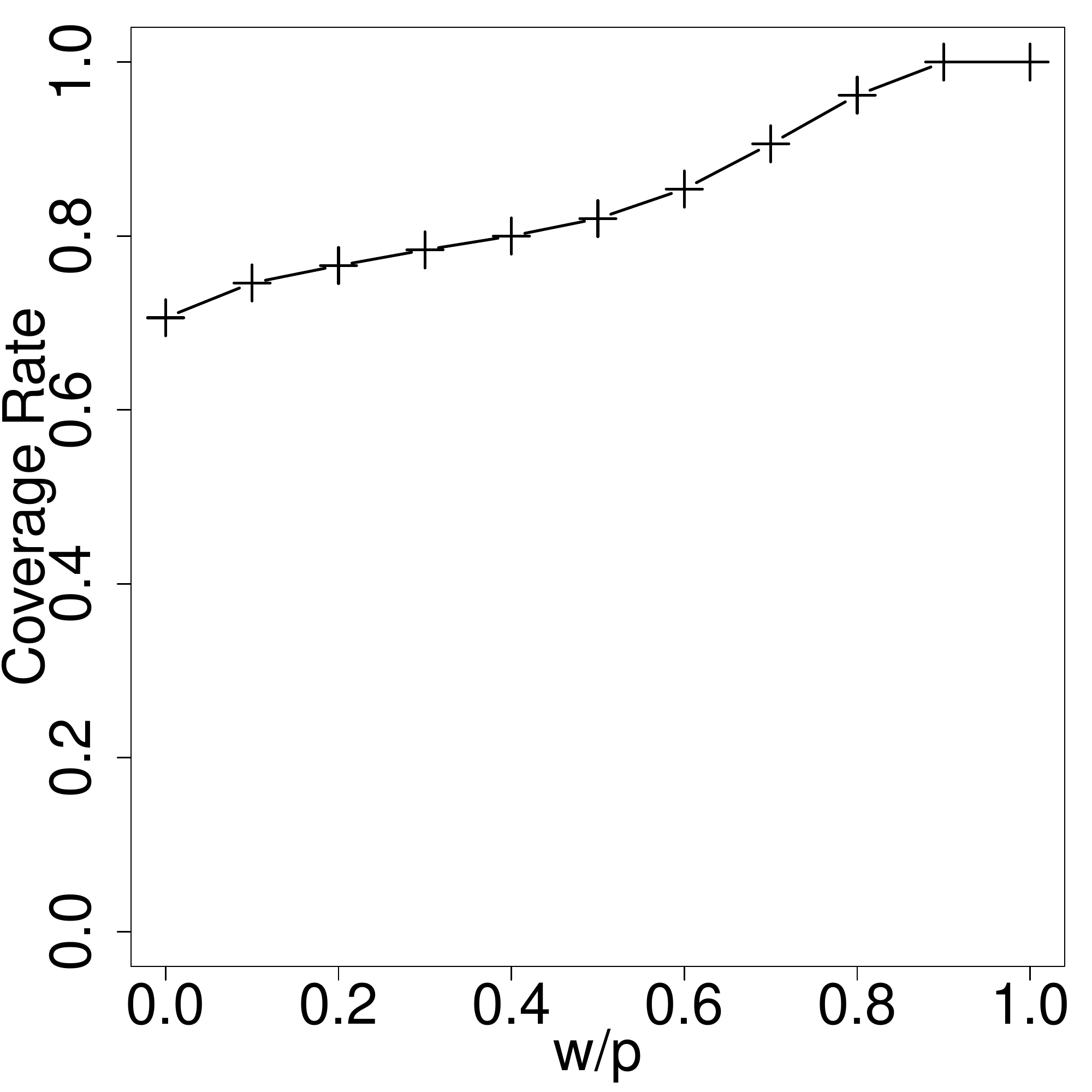}
}
\hspace{0 in} 
\subfigure[$p=200$]{
\includegraphics[width=0.3\textwidth]{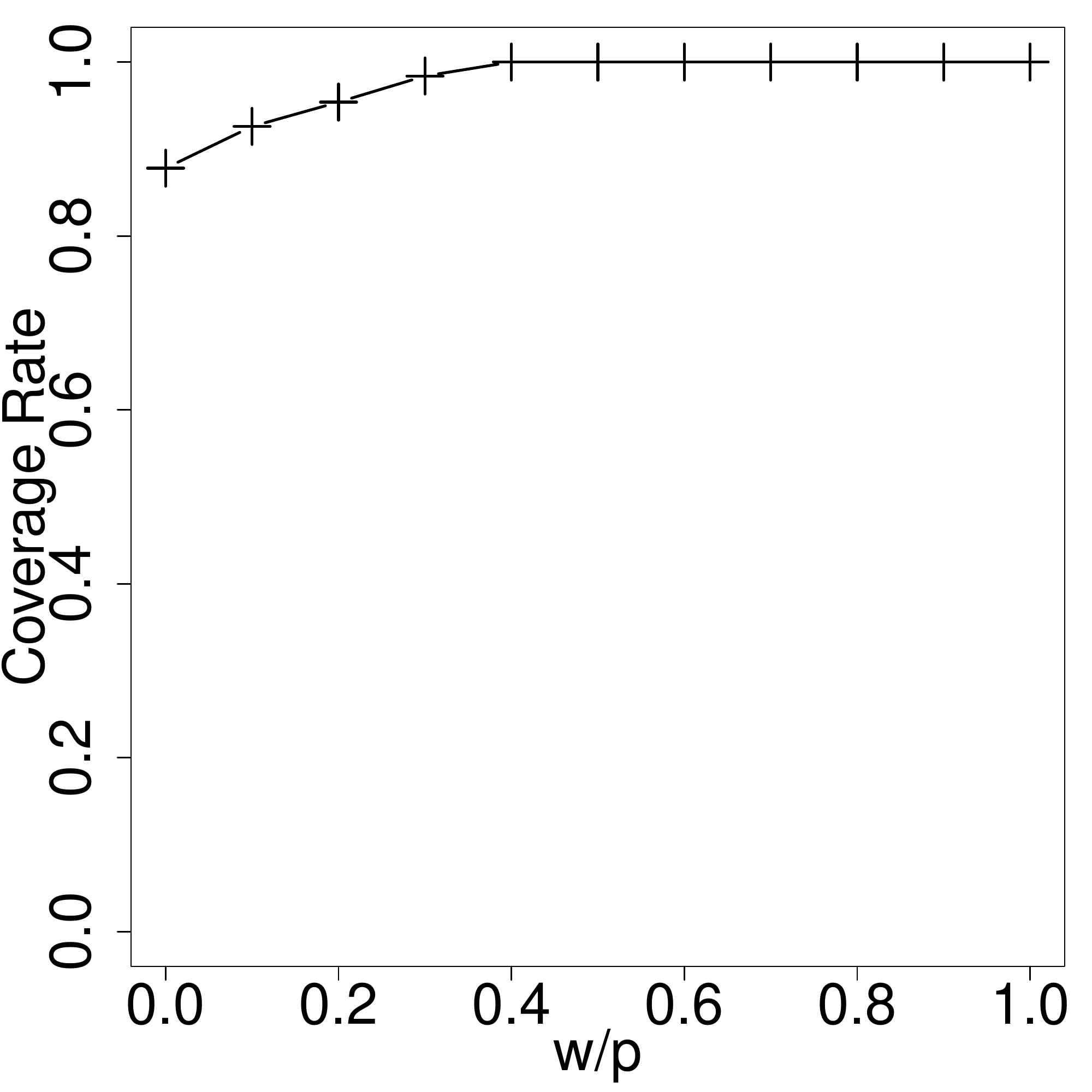}
}
\caption{
MUCs of Algorithms 1 and 2 for different scenarios with independent covariates.
The top row shows the results of Algorithms 1 and 2:
(a) $p=8$, $p^*=3$; (b) $p=10$, $p^*=4$; (c) $p=15$, $p^*=6$.
The second row shows only Algorithm 2 because Algorithm 1 is infeasible in 
these cases: (d) 
$p=50$, $p^*=8$; (e) $p=100$, $p^*=10$; (f) 
$p=200$, $p^*=12$.
The Adaptive Lasso is used as the variable selection method.
}
\label{fig: illu}
\end{figure}

We consider six scenarios: (a) $p=8$, $p^*=3$; 
(b) $p= 10$, $p^*=4$; (c) $p=15$, $p^*=6$; (d) $p=50$, $p^*=8$; (e) $p=100$, 
$p^*=10$; and (f) $p=200$, $p^*=12$.
We set $n=300$, $B=1000$, $\theta_j=1$ and $\sigma=1$.  
In addition, we let $\rho=0$ and $0.5$ to simulate the cases of independent and 
correlated covariates.
Adaptive Lasso with 10-fold cross-validation is used as the model selection 
method.
Through simulation, we see that the MUCs from Algorithms 1 
and 2 are very similar under these six scenarios using both independent and 
correlated covariates, 
indicating the performance (in terms of coverage) of Algorithms 1 and 2 are 
almost the 
same. 
For details of these MUCs, please see Figure~\ref{fig: illu} and Web Appendix A Figure~\ref{fig:illu_correlated}.
Note that Algorithm 1 uses the exhaustive search so its MUC is the highest 
possible and cannot be improved.
Therefore, we conclude that Algorithm 2 performs (nearly) optimally as well.
However, using Algorithm 1, MCB becomes impossible to obtain when $p$ is 
large because it requires too much time, while using Algorithm 2, we can easily 
obtain MCB.   
We further explore their computational times.
To complete one MC iteration, 
Algorithm 1 takes 22.37, 207.64, 67750.81 seconds in scenarios (a), (b), and (c), and 
cannot compute at all for larger $p$s in scenarios (d), (e), and (f).
On the other hand, Algorithm 2 takes less than 1 second in scenarios (a), (b), 
and (c), and 5.01, 22.70, 201.43 seconds in scenarios (d), (e), and (f).
Such a phenomenon is not a surprise as explained in Section 
\ref{sec:algorithms}.
Therefore, we adopt Algorithm 2 throughout the rest of the simulation 
studies.

\subsection{Assessing Model Selection Uncertainty}\label{sec:comp_model_sel}

One of the greatest advantages of MCB is to provide a platform to assess 
uncertainties of different model selection methods.
We use the proposed method to demonstrate such an advantage under linear 
regressions.

\begin{figure}
\centering
\subfigure[Normal distribution, $\rho=0$]{
\includegraphics[width=0.45\textwidth]{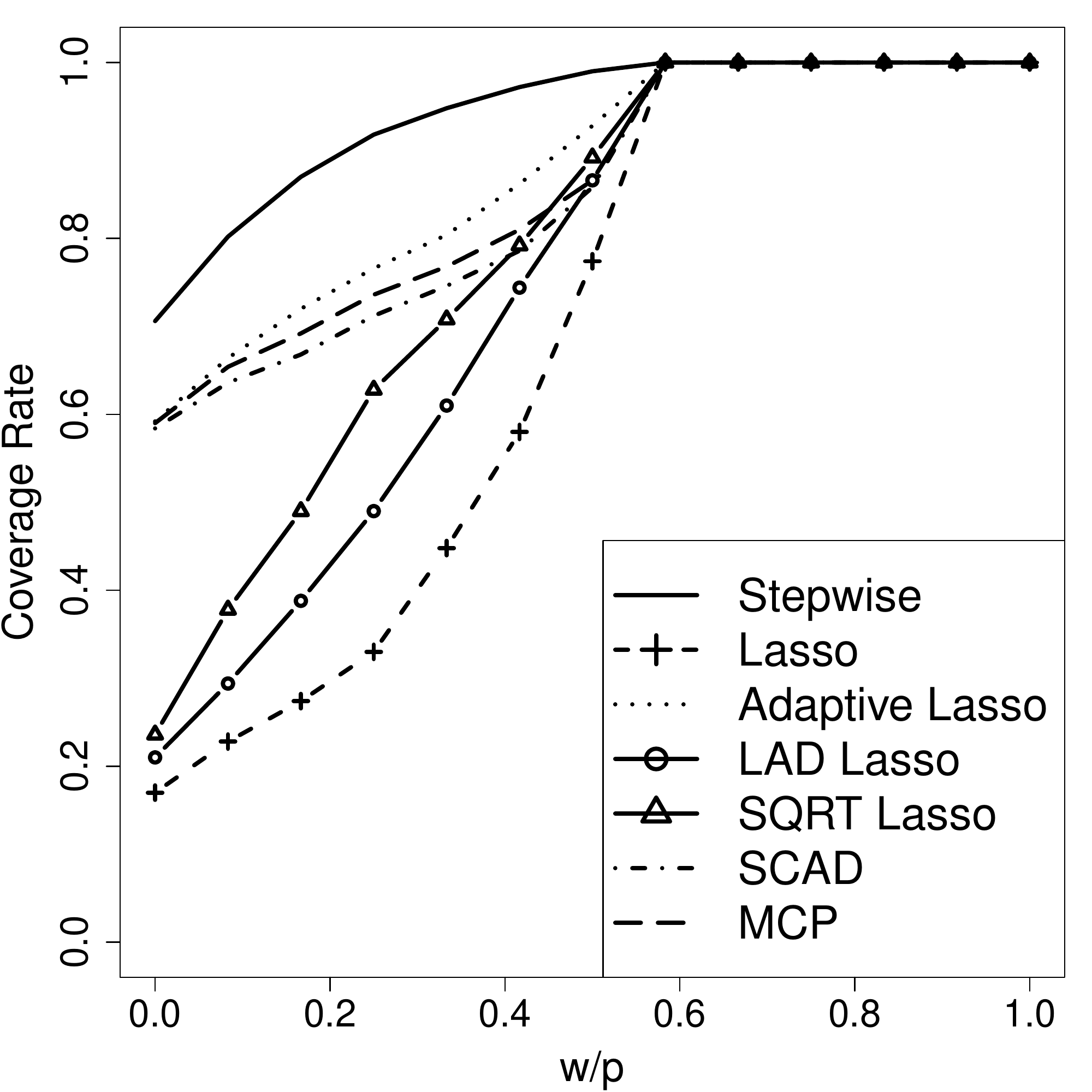}}
\hspace{0in} 
\subfigure[Laplace distribution, $\rho=0$]{
\includegraphics[width=0.45\textwidth]{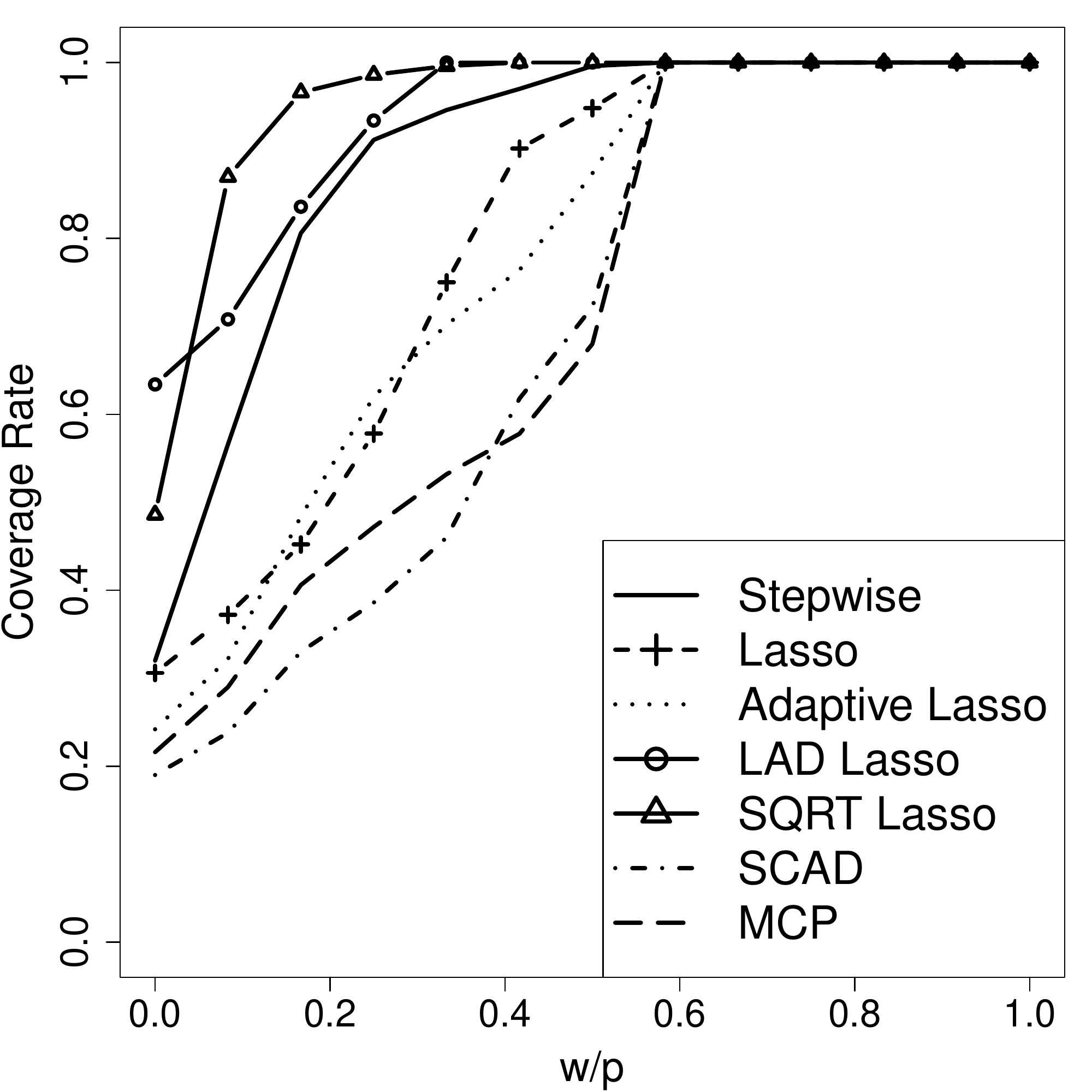}}
\subfigure[Normal distribution, $\rho=0.5$]{
\includegraphics[width=0.45\textwidth]{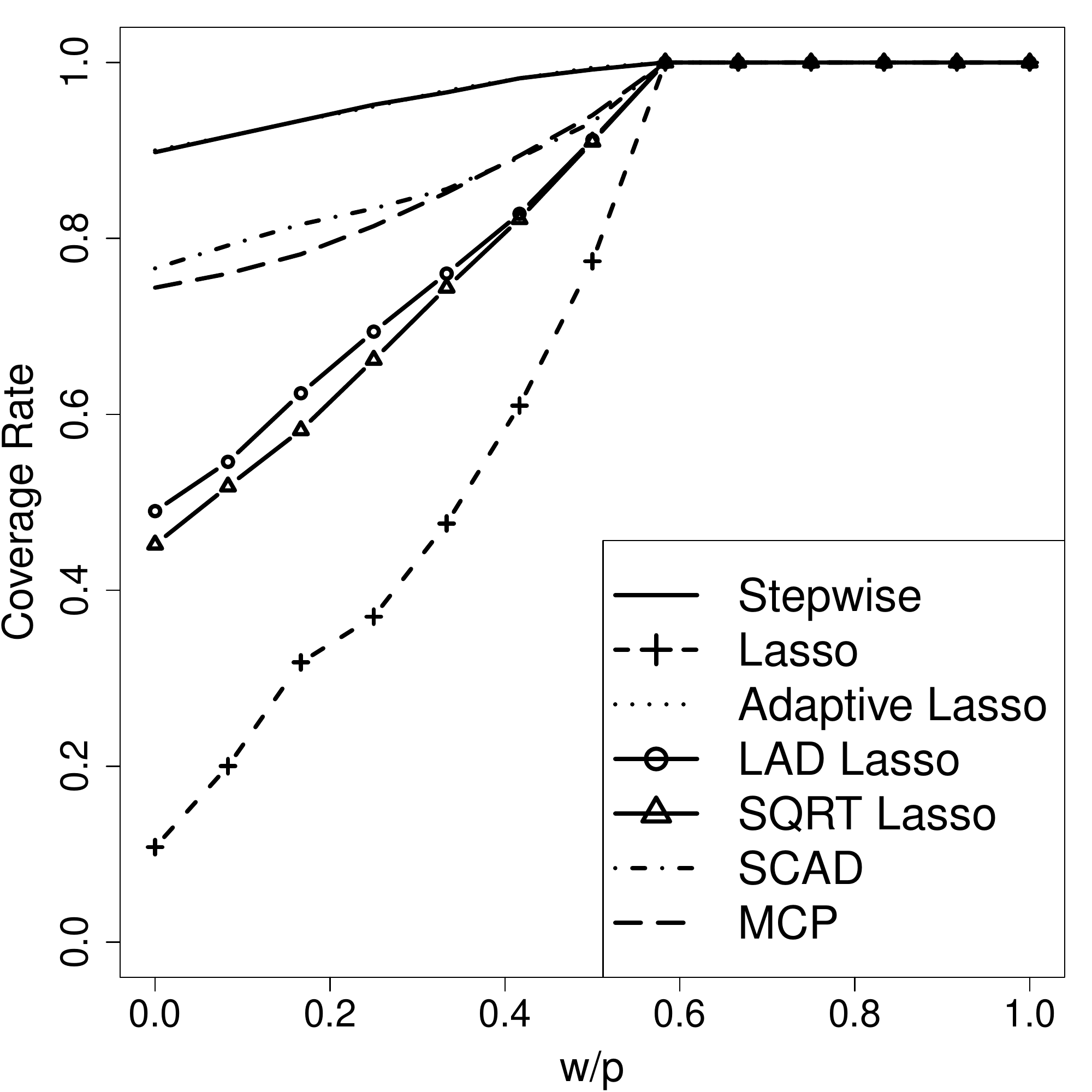}}
\hspace{0in} 
\subfigure[Laplace distribution, $\rho=0.5$]{
\includegraphics[width=0.45\textwidth]{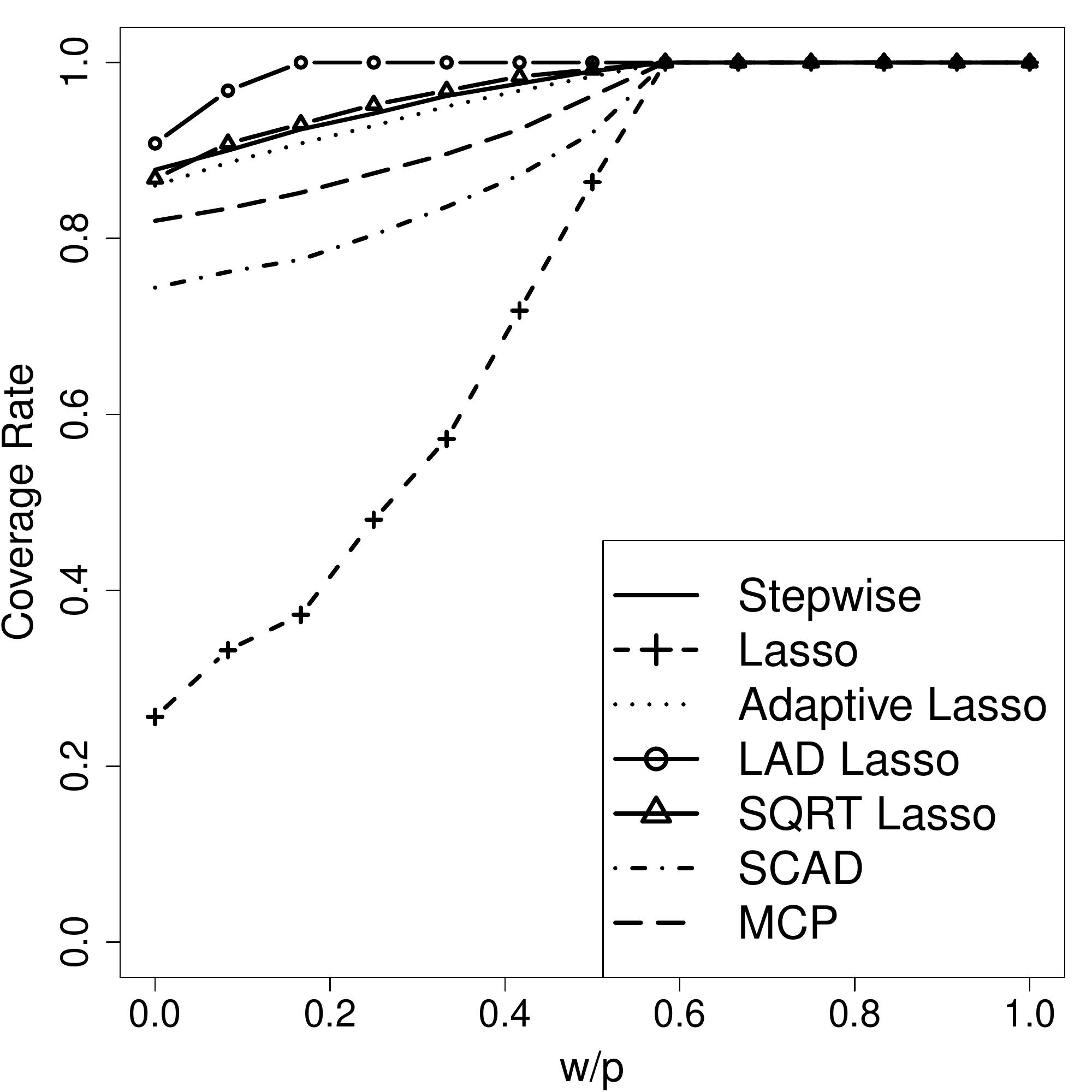}}
\caption{MUCs of different model selection methods for linear regression. 
The data generating model 
is $ y=\sum_{j=1}^{5} 1 \times x_j+\sum_{j=6}^{12}0 \times x_j +\epsilon$.   
Panel: (a) 
$\epsilon\sim N(0,1)$, (b) $\epsilon \sim 
\textup{Laplace}(0,\sqrt{1/2})$. $n=300$ and $ B=1000$.   
Stepwise regression using BIC, Lasso, Adaptive Lasso, LAD Lasso, SQRT Lasso, 
SCAD and MCP are compared.
Tuning parameters are chosen based on 10-fold cross-validation.}
\label{fig:varsel} 
\end{figure}

We generate data according to the linear regression model and set $p^*=5$, 
$p=12$, $\theta_j=1$, $n=300$ and $B=1000$.  
We simulate the random error under two scenarios: $\epsilon_i  \sim N(0,1)$ and 
$\epsilon_i \sim \textup{Laplace}(0,\sqrt{1/2})$, which give $\epsilon_i$ the 
same variance.
In addition, we let $\rho=0$ and $0.5$ to simulate the cases of independent and 
correlated covariates.
We compare various model selection methods with different loss functions and 
penalty functions.
In particular, we include stepwise regression using BIC, Lasso, Adaptive Lasso, 
LAD Lasso, SQRT Lasso, SCAD and MCP.
Note that Lasso, Adaptive Lasso, SCAD, and MCP have the same loss function, 
while Lasso, LAD Lasso, SQRT Lasso have the same penalty. 
Figure~\ref{fig:varsel} shows the MUCs of different model selection methods 
under these simulation settings.
Note that a more stable model selection method will have its MUC arch further 
towards the upper left corner.
For the normal distribution, stepwise, Adaptive Lasso, SCAD and MCP perform 
better than the rest, because their loss function is compatible with the normal 
random errors.
For the Laplace distribution, LAD Lasso and SQRT Lasso perform better than the 
rest, because LAD Lasso and SQRT Lasso are more robust against heavy 
tail random errors, therefore, offer lower model selection uncertainty.
Such a phenomenon is consistent for both independent and correlated covariates.
For numerical studies, stepwise regression is available in \texttt{leaps}, 
Lasso is available in \texttt{glmnet}, 
Adaptive Lasso is available in \texttt{parcor}, 
LAD Lasso and SQRT lasso are available in \texttt{flare}, 
and MCP and SCAD are available in \texttt{ncvreg}.

\subsection{Comparison of MCB and VSCS}\label{sec:comp_MCB_VSCS}

We further compare the performance of MCB with VSCS
\protect\citep{ferrari2015confidence}.  
Note that MCB is equipped with LBM and UBM and all the models contained in MCB 
are nested between LBM and UBM,
whereas VSCS contains a basket of models which are the survivors of the 
F-test and hence are not necessarily nested.
MCB is implemented by Adaptive Lasso while VSCS is implemented by F-test.

We simulate data according to the linear 
regression model with $n=100$, $B=1000$, $p^*=5$, $p=10$, $\theta_j=\gamma^j$ 
and $\sigma=1$. 
We set $\rho=0$, 0.25 and 0.5 to simulate the cases of independent and 
correlated 
covariates, and further set $\gamma=1$ and $0.6$ to 
simulate the constant coefficients and power decaying coefficients. 
Under these scenarios, we present the true model coverage rates of MCB and 
VSCS, cardinalities (i.e., the number of unique models returned by MCB and 
VSCS) in Table~\ref{tab:MCBVSCS}.  
As we can see, under the same confidence levels, both MCB and VSCS have 
approximately the same coverage rates, suggesting both of them are valid tools.
MCB occasionally has higher coverage rates because of the design of the 
algorithm.  
On the other hand, MCB consists fewer models than VSCS and also preserves 
a structure of its model composition, which means MCB is more efficient in 
capturing the true model and its results are easier to interpret.
Such a phenomenon persists even when we have correlated covariates and 
power decaying coefficients (i.e., $\rho=0.5$ and $\gamma=0.6$).
However, using the power decaying coefficients, 
it is much harder to select the models. 
Therefore, the cardinality increases for both MCB and VSCS,
but MCB still hold advantage over VSCS.
The advantage of MCB over VSCS in these scenarios is partially 
due to the fact that VSCS does not incorporate an UBM 
(or equivalently, VSCS has the UBM of the full model since the full model always survives the F-test).
On the other hand, MCB has both LBM and UBM, therefore, contains fewer 
models than VSCS.
Lastly, VSCS uses F-tests to select models whereas MCB uses 
Adaptive Lasso, which also accounts for the advantage of MCB.

\section{Conclusions}\label{sec:conc}

This paper introduces the concept of model confidence bounds (MCB) for 
variable selection and proposes an efficient algorithm to obtain MCB.
Rather than blindly relying on a single selected model without knowing its 
credibility, MCB yields two bounds for models that capture a 
group of nested models which contains the true model at a given confidence 
level;
it extends the notion of confidence interval for population parameter to the 
variable selection problem.
MCB can be used as a model selection diagnostic tool as well as a platform for 
assessing different model selection methods' uncertainty.
By comparing the proposed MUC, we can evaluate the stabilities of model 
selection methods, just like we use confidence intervals to evaluate estimators.
Therefore, MCB provide more insights into the existing variable selection 
methods and a deeper understanding of observed data sets.

There are many directions remaining for further research.
For example, throughout this paper, we have assumed that MCB consists of one
LBM and one UBM.
However, it may be possible to find multiple LBMs that are statistically
equivalent \protect\citep{ferrari2015confidence}.
Thus, we may consider a richer structure of MCB with multiple LBMs or even
multiple UBMs.
In addition, MCB can be extended to other classes models, such as, GLM and time series models \protect\citep{meier2008group, Zhou2015}.
We leave these topics for future research.

\bibliographystyle{biom.bst}
\bibliography{Bibliography-cs}

\newpage

\begin{landscape}
 \begin{table}
 \centering
 \footnotesize
 \begin{tabular}{cccccccccccccccc}
 \hline
 Method & $1-\alpha$ & MCB & \texttt{bmi} & \texttt{ltg} & \texttt{map} 
 & 
 \texttt{tc } & \texttt{sex} & \texttt{tch} & \texttt{ldl} & \texttt{glu} & 
 \texttt{hdl} & \texttt{age} & Width & Cardinality & CR\\
 \hline                       
 \multirow{4}{*}{MCB ALasso} & \multirow{2}{*}{$0.95$} & UBM & 
 \cellcolor{gray!90}  
 &\cellcolor{gray!90}&\cellcolor{gray!90}&\cellcolor{gray!90}&\cellcolor{gray!90}&\cellcolor{gray!90}&\cellcolor{gray!90}&\cellcolor{gray!90}&\cellcolor{gray!90}&
  
 &\multirow{2}{*}{6}&\multirow{2}{*}{64}&\multirow{2}{*}{0.975}\\ 
 & & LBM & \cellcolor{gray!30}&\cellcolor{gray!30}&\cellcolor{gray!30}&
                        & & & & & & & & &\vspace{0.8em}\\ 

 &\multirow{2}{*}{$0.75$}&UBM&\cellcolor{gray!90}&\cellcolor{gray!90}
 &\cellcolor{gray!90}&\cellcolor{gray!90}&\cellcolor{gray!90}
 &\cellcolor{gray!90}&\cellcolor{gray!90}&\cellcolor{gray!90}&
 &&\multirow{2}{*}{4}&\multirow{2}{*}{16}&\multirow{2}{*}{0.811}\\
                       &  &LBM&\cellcolor{gray!30}&\cellcolor{gray!30}
                       &\cellcolor{gray!30}&\cellcolor{gray!30}& & & & & & & & 
                       &\vspace{0.8em}\\ 
 \multirow{4}{*}{MCB Lasso}& \multirow{2}{*}{$0.95$}&UBM&\cellcolor{gray!90}
 &\cellcolor{gray!90}&\cellcolor{gray!90}&\cellcolor{gray!90}&\cellcolor{gray!90}
 &\cellcolor{gray!90}&\cellcolor{gray!90}&\cellcolor{gray!90}&\cellcolor{gray!90}
 & &\multirow{2}{*}{5}&\multirow{2}{*}{32}&\multirow{2}{*}{0.999}\\
                       & &LBM&\cellcolor{gray!30}&\cellcolor{gray!30}
                       &\cellcolor{gray!30}&\cellcolor{gray!30}& & & & & & & & 
                       &\vspace{0.8em}\\                   
                       
 &\multirow{2}{*}{$0.75$}&UBM&\cellcolor{gray!90}&\cellcolor{gray!90}
 &\cellcolor{gray!90}&\cellcolor{gray!90}&\cellcolor{gray!90}&\cellcolor{gray!90}
 &\cellcolor{gray!90}&\cellcolor{gray!90}&\cellcolor{gray!90}
 & &\multirow{2}{*}{4}&\multirow{2}{*}{16}&\multirow{2}{*}{0.805}\\
                     &   &LBM&\cellcolor{gray!30}&\cellcolor{gray!30}
                     &\cellcolor{gray!30}&\cellcolor{gray!30}&\cellcolor{gray!30}
                     & & & & & & & &\vspace{0.8em}\\ 
                     
 \multirow{4}{*}{MCB 
 Stepwise}&\multirow{2}{*}{$0.95$}&UBM&\cellcolor{gray!90}
 &\cellcolor{gray!90}&\cellcolor{gray!90}&\cellcolor{gray!90}
 &\cellcolor{gray!90}&\cellcolor{gray!90}&\cellcolor{gray!90}
 &\cellcolor{gray!90}&\cellcolor{gray!90}
 & &\multirow{2}{*}{6}&\multirow{2}{*}{64}&\multirow{2}{*}{0.973}\\
                       & &LBM&\cellcolor{gray!30}&\cellcolor{gray!30}
                       &\cellcolor{gray!30}& & & & & & & & & 
                       &\vspace{0.8em}\\                   
 &\multirow{2}{*}{$0.75$}&UBM&\cellcolor{gray!90}&\cellcolor{gray!90}
 &\cellcolor{gray!90}&\cellcolor{gray!90}&\cellcolor{gray!90}&\cellcolor{gray!90}
 &\cellcolor{gray!90}&\cellcolor{gray!90}
 & & &\multirow{2}{*}{5}&\multirow{2}{*}{32}&\multirow{2}{*}{0.812}\\
                       & 
                       &LBM&\cellcolor{gray!30}&\cellcolor{gray!30}&\cellcolor{gray!30}
                       &&&&&&&&&&\vspace{0.4em}\\               
 \hline
 ALasso& & &\cellcolor{gray!60} &\cellcolor{gray!60}
 &\cellcolor{gray!60} & \cellcolor{gray!60} & \cellcolor{gray!60} 
 &\cellcolor{gray!60} & \cellcolor{gray!60} & & & & - & - & - \\
 Lasso   &  & &\cellcolor{gray!60}&\cellcolor{gray!60}&\cellcolor{gray!60}
 &\cellcolor{gray!60}&\cellcolor{gray!60}&\cellcolor{gray!60}
 &  &\cellcolor{gray!60}&\cellcolor{gray!60}& & - & - & -\\
 Stepwise& & & \cellcolor{gray!60}&\cellcolor{gray!60}&\cellcolor{gray!60}
 & &\cellcolor{gray!60}& & & & \cellcolor{gray!60}& & - & - & - 
 \vspace{0.4em}\\
 \hline
 
 & & UBM & \cellcolor{gray!90}&\cellcolor{gray!90}
 &\cellcolor{gray!90}&\cellcolor{gray!90}&\cellcolor{gray!90}&\cellcolor{gray!90}
 &\cellcolor{gray!90}&\cellcolor{gray!90}&\cellcolor{gray!90}
 &\cellcolor{gray!90} & & & \\
  
 \multirow{7}{*}{VSCS}& \multirow{7}{*}{$0.95$} & LBM1 &   &   & 
 \cellcolor{gray!30} & 
 \cellcolor{gray!30} & 
 \cellcolor{gray!30} &   & \cellcolor{gray!30} &   & \cellcolor{gray!30} & & 
 \multirow{7}{*}{-} & 
 \multirow{7}{*}{528} 
 &\multirow{7}{*}{-}\\
 
 & & LBM2 & \cellcolor{gray!30} &   & \cellcolor{gray!30} &   &   &   
 &   &   &   & & & & \\
 
 & & LBM3 & \cellcolor{gray!30} &   &   &   &   &   &   &   & 
 \cellcolor{gray!30} & & & &\\
 
 & & LBM4 & \cellcolor{gray!30} &   &   &   &   & \cellcolor{gray!30} 
 &   &   &   & & & &\\
 
 & & LBM5 & \cellcolor{gray!30} & \cellcolor{gray!30} &   &   &   &   
 &   &   &   & & & &\\
 
 & & LBM6 & \cellcolor{gray!30} &   &   &   &   &   &   & 
 \cellcolor{gray!30} &   & & & &\\
 
 & & LBM7 &   & \cellcolor{gray!30} & \cellcolor{gray!30} &   &   &   
 &   &   &   & & & &\vspace{0.8em}\\

 & & UBM & \cellcolor{gray!90}&\cellcolor{gray!90}
 &\cellcolor{gray!90}&\cellcolor{gray!90}&\cellcolor{gray!90}&\cellcolor{gray!90}
 &\cellcolor{gray!90}&\cellcolor{gray!90}&\cellcolor{gray!90}
 &\cellcolor{gray!90} & & & \\

 \multirow{4}{*}{VSCS}& \multirow{4}{*}{0.75} & LBM1 & \cellcolor{gray!30} &   
 & 
 \cellcolor{gray!30} &   &   &   
 &   &   & \cellcolor{gray!30} & & \multirow{4}{*}{-} & \multirow{4}{*}{288} & 
 \multirow{4}{*}{-}\\
 
 & & LBM2 & \cellcolor{gray!30} &   & \cellcolor{gray!30} &   &   & 
 \cellcolor{gray!30} &   &   &   & & & &\\
 
 & & LBM3 & \cellcolor{gray!30} &   &   & \cellcolor{gray!30} &   &   & 
 \cellcolor{gray!30} &   & \cellcolor{gray!30} & & & &\\
 & & LBM4 & \cellcolor{gray!30} & \cellcolor{gray!30} &   &   &   &   
 &   &   &   & & & &\vspace{0.4em}\\
 \hline
 \end{tabular}
 \caption{Results of MCB with different variable selection methods (Adaptive 
 Lasso, Lasso, and stepwise regression using BIC) and VSCS at 75\% and 95\% 
 confidence levels. 
 Light and dark gray cells denote that the predictors are in the LBM or UBM. 
 $B=1000$ bootstrap samples are generated.  
 Medium gray cells in the middle three rows denote that the predictors are 
 selected in the single model after applying these variable selection methods 
 to the original dataset.
 The bottom rows indicate the results of VSCS. 
 Note that VSCS has multiple LBM at each confidence level and does not have 
 UBMs.
 In addition, VSCS does not have width or CR.
 }\label{tab:MCB}
 \end{table}
\end{landscape}

\begin{table}
\centering
\footnotesize
\begin{tabular}{ccccccc}
\hline
&&& \multicolumn{2} {c} {MCB} & \multicolumn{2} {c} {VSCS} \\
$\rho$ & $\gamma$ & $100(1-\alpha)\%$& Coverage Rate & Cardinality & Coverage 
Rate & Cardinality\\
\hline                       
0&		1&	95\%&	0.93&	31.65&	0.94&	32.57\\
0&		1&	90\%&	0.89&	10.09&	0.88&	29.69\\
0&		1&	85\%&	0.87&	5.51&	0.82&	27.58\\
0&		1&	80\%&	0.85&	3.99&	0.77&	25.74\\
0&		1&	75\%&	0.83&	3.14&	0.72&	23.91\\
0&		1&	70\%&	0.79&	2.60&	0.67&	22.25\\
0&		1&	65\%&	0.76&	2.27&	0.62&	20.53\\
0&		1&	60\%&	0.74&	1.98&	0.57&	18.95\\
0.25&	1&	95\%&	0.96&	8.02&	0.96&	36.36\\
0.25&	1&	90\%&	0.93&	4.13&	0.92&	31.96\\
0.25&	1&	85\%&	0.90&	3.06&	0.85&	29.17\\
0.25&	1&	80\%&	0.88&	2.58&	0.79&	27.01\\
0.25&	1&	75\%&	0.86&	2.20&	0.76&	25.15\\
0.25&	1&	70\%&	0.82&	1.94&	0.71&	23.38\\
0.25&	1&	65\%&	0.80&	1.73&	0.68&	21.65\\
0.25&	1&	60\%&	0.77&	1.60&	0.62&	19.94\\
0.5&	1&	95\%&	0.92&	8.25&	0.95&	49.74\\
0.5&	1&	90\%&	0.88&	4.42&	0.89&	39.95\\
0.5&	1&	85\%&	0.84&	3.21&	0.85&	34.60\\
0.5&	1&	80\%&	0.81&	2.64&	0.80&	30.85\\
0.5&	1&	75\%&	0.80&	2.33&	0.76&	28.11\\
0.5&	1&	70\%&	0.77&	2.02&	0.71&	25.56\\
0.5&	1&	65\%&	0.74&	1.80&	0.66&	23.42\\
0.5&	1&	60\%&	0.73&	1.63&	0.62&	21.34\\
0&		0.6&	95\%&	0.98&	266.20&	0.94&	291.80\\
0&		0.6&	90\%&	0.95&	218.20&	0.91&	237.50\\
0&		0.6&	85\%&	0.93&	190.70&	0.84&	203.50\\
0&		0.6&	80\%&	0.85&	162.90&	0.81&	176.60\\
0&		0.6&	75\%&	0.77&	140.50&	0.75&	155.60\\
0&		0.6&	70\%&	0.71&	121.00&	0.70&	137.00\\
0&		0.6&	65\%&	0.69&	113.30&	0.64&	121.10\\
0&		0.6&	60\%&	0.64&	90.20&	0.61&	106.00\\
0.5&	0.6&	95\%&	0.94&	279.00&	0.92&	320.20\\
0.5&	0.6&	90\%&	0.90&	214.40&	0.87&	262.50\\
0.5&	0.6&	85\%&	0.84&	176.60&	0.84&	227.30\\
0.5&	0.6&	80\%&	0.81&	156.80&	0.82&	201.60\\
0.5&	0.6&	75\%&	0.73&	132.50&	0.79&	179.60\\
0.5&	0.6&	70\%&	0.67&	115.50&	0.71&	160.80\\
0.5&	0.6&	65\%&	0.61&	99.80&	0.66&	143.50\\
0.5&	0.6&	60\%&	0.51&	88.00&	0.60&	127.90\\

\hline
\end{tabular}
\caption{Comparison of MCB and VSCS in terms of cardinalities and coverage 
rates at various confidence levels using both independent and correlated 
covariates and constant and exponentially decaying coefficients.}
\label{tab:MCBVSCS}
\end{table}

\end{document}